\PassOptionsToPackage{prologue,dvipsnames}{xcolor}

\documentclass[acmsmall,screen]{acmart}
\usepackage{colortbl}
\usepackage{enumitem}
\usepackage{listings} 
\usepackage{xcolor}
\usepackage[ruled,linesnumbered,noend]{algorithm2e}
\usepackage{makecell}
\usepackage{multirow}
\usepackage{colortbl}
\usepackage{threeparttable}
\usepackage{subfigure}
\usepackage{xspace}
\usepackage{float}
\usepackage{graphicx}
\usepackage{textcomp}
\usepackage{diagbox}
\usepackage{rotating}
\usepackage{placeins}
\usepackage{url}
\usepackage{hyperref}
\usepackage{wrapfig}
\usepackage{caption}
\usepackage{tikz}
\definecolor{mybrown}{RGB}{184, 80, 5}
\definecolor{bg}{rgb}{0.95,0.95,0.95}
\newcommand{\colortexttt}[2]{\textcolor{#1}{\texttt{#2}}}

\newcommand{\hollowcirc}{%
    \begin{tikzpicture}[baseline=-0.5ex]
        \draw (0,0) circle (2.5pt); 
    \end{tikzpicture}%
}

\newcommand{\solidcirc}{%
    \begin{tikzpicture}[baseline=-0.5ex]
        \fill[black] (0,0) circle (2.5pt); 
    \end{tikzpicture}%
}

\newcommand{\halfsolidcirc}{%
    \begin{tikzpicture}[baseline=-0.5ex]
        \draw (0,0) circle (2.5pt);
        \begin{scope}
            \clip (-2.5pt,-2.5pt) rectangle (0pt,2.5pt);
            \fill[black] (0,0) circle (2.5pt);
        \end{scope}
    \end{tikzpicture}%
}

\newcommand{\threefourhollowcirc}{%
    \begin{tikzpicture}[baseline=-0.5ex]
        \draw (0,0) circle (2.5pt); 
        \begin{scope}
            \clip (-2.5pt,-2.5pt) rectangle (-1.25pt,2.5pt); 
            \fill[black] (0,0) circle (2.5pt);
        \end{scope}
    \end{tikzpicture}%
}

\graphicspath{{figs/}}
\AtBeginDocument{%
  }

\setcopyright{acmlicensed}
\copyrightyear{2025}
\acmYear{2025}
\acmDOI{XXXXXXX.XXXXXXX}

\acmJournal{JACM}
\acmVolume{37}
\acmNumber{4}
\acmMonth{8}

\newcommand{\rustc}{\texttt{rustc}\xspace}

\usepackage{pifont}
\newcommand{\tmark}{{\scriptsize\ding{228}}}

\begin{document}


\title[{An Empirical Study of Rust-Specific Bugs in the \texttt{rustc} Compiler}]{An Empirical Study of Rust-Specific Bugs \\ in the \texttt{rustc} Compiler}

\author{Zixi Liu}
\email{zxliu@smail.nju.edu.cn}
\affiliation{%
  \institution{State Key Laboratory for Novel Software Technology Nanjing University}
  \city{Nanjing}
  \country{China}
}

\author{Yang Feng}
\email{fengyang@nju.edu.cn}
\affiliation{%
  \institution{State Key Laboratory for Novel Software Technology Nanjing University}
  \city{Nanjing}
  \country{China}
}

\author{Yunbo Ni}
\email{yunboni@smail.nju.edu.cn}
\affiliation{%
  \institution{State Key Laboratory for Novel Software Technology Nanjing University}
  \city{Nanjing}
  \country{China}
}

\author{Shaohua Li}
\email{shaohuali@cse.cuhk.edu.hk}
\affiliation{%
  \institution{The Chinese University of Hong Kong}
  \country{China}
}

\author{Xizhe Yin}
\email{xizheyin@smail.nju.edu.cn}
\affiliation{%
  \institution{State Key Laboratory for Novel Software Technology Nanjing University}
  \city{Nanjing}
  \country{China}
}

\author{Qingkai Shi}
\email{qingkaishi@nju.edu.cn}
\affiliation{%
  \institution{State Key Laboratory for Novel Software Technology Nanjing University}
  \city{Nanjing}
  \country{China}
}

\author{Baowen Xu}
\email{bwxu@nju.edu.cn}
\affiliation{%
  \institution{State Key Laboratory for Novel Software Technology Nanjing University}
  \city{Nanjing}
  \country{China}
}

\author{Zhendong Su}
\email{zhendong.su@inf.ethz.ch}
\affiliation{%
  \institution{ETH Zurich}
  \country{Switzerland}
}

\renewcommand{\shortauthors}{Liu et al.}

\begin{abstract}
Rust is gaining popularity for its well-known memory safety guarantees and high performance, distinguishing it from C/C++ and JVM-based languages.
Its compiler, \rustc, enforces these guarantees through specialized mechanisms such as trait solving, borrow checking, and specific optimizations.
However, Rust's unique language mechanisms introduce complexity to its compiler, leading to Rust-specific compiler bugs that are less common in traditional compilers.
With Rust's increasing adoption in safety-critical domains, understanding these language mechanisms and their impact on compiler bugs is essential for improving the reliability of both \rustc and Rust programs.
Yet, we still lack a large-scale, detailed, and in-depth study of Rust-specific bugs in \rustc.

To bridge this gap, this work conducts a comprehensive and systematic study of Rust-specific bugs in \rustc, with a particular focus on the components that support its unique language features.
Our analysis examines issues and fixes reported between 2022 and 2024, with a manual review of 301 valid issues. We categorize these bugs based on their causes, symptoms, affected compilation stages, and test case characteristics. 
Additionally, we evaluate existing \rustc testing tools to assess their effectiveness and limitations.
Our key findings include: 
(1) \rustc bugs primarily arise from Rust's type system and lifetime model, with frequent errors in the High-Level Intermediate Representation (HIR) and Mid-Level Intermediate Representation (MIR) modules due to complex checkers and optimizations; 
(2) bug-revealing test cases often involve unstable features, advanced trait usages, lifetime annotations, standard APIs, and specific optimization levels;
(3) while both valid and invalid programs can trigger bugs, existing testing tools struggle to detect non-crash errors, underscoring the need for further advancements in \rustc testing.
\end{abstract}




\received{20 February 2007}
\received[revised]{12 March 2009}
\received[accepted]{5 June 2009}

\maketitle

\section{Introduction}\label{sec:intro}
As the demand for more secure programming paradigms grows, the need for languages with fewer memory vulnerabilities becomes more recognized.
For instance, United States White House recently emphasized the importance of adopting memory-safe languages, with Rust recognized as a leading example~\cite{WhiteHou52:online}.
Rust's unique principles, such as ownership, borrowing, and lifetimes, enable developers to write both secure and efficient code. 
Additionally, Rust's focus on zero-cost abstractions and fearless concurrency has made it particularly popular in system programming~\cite{jung2021safe, klabnik2023rust}. 
Recently, there is an increasing trend to re-engineer widely used software systems in Rust~\cite{Servothe7:online, RedoxYou24:online, TiKVTiKV64:online, StratisS70:online, Cloudfla15:online}.

\begin{figure}[htbp]
  \centering
  \includegraphics[width = 0.9\linewidth]{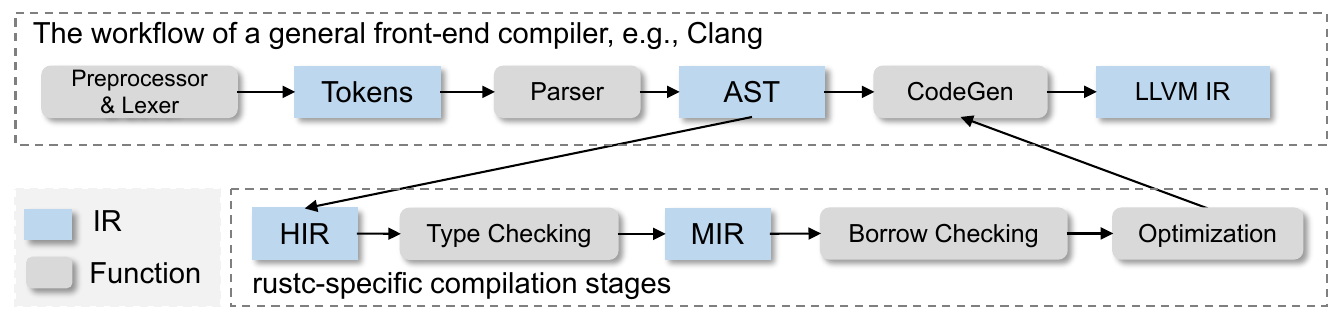}
  \caption{The high-level workflow of \rustc and a general front-end compiler.}
  \vspace{-10pt}
  \label{fig:rustc_workflow}
\end{figure}

The primary compiler for Rust is the official open-source \rustc~\cite{Whatisru31:online}, which is written in Rust and uses LLVM~\cite{lattner2004llvm} as its back-end. 
Similar to classic front-end compilers, such as Clang, \rustc translates source code into LLVM IR, but it introduces additional intermediate representations (IRs) and compilation components to support Rust's unique language mechanisms. 
As illustrated in Figure~\ref{fig:rustc_workflow}, \rustc follows a multi-stage compilation workflow tailored to enforce Rust's strict safety guarantees and advanced type system.
After parsing the input program, the Abstract Syntax Tree (AST) is transformed into the High-Level Intermediate Representation (HIR), which abstracts over syntactic details to facilitate type inference, type checking, and trait resolution. 
This processing is crucial yet complex due to Rust's trait system, which enables zero-cost abstractions while supporting highly flexible usage patterns.
Additionally, some data types are annotated with lifetimes, posing challenges for \rustc's type inference.
Then, the HIR is lowered to Mid-Level Intermediate Representation (MIR), a control-flow-oriented representation crucial for enforcing Rust's ownership model, borrow checking, and move semantics. 
Before generating LLVM IR, \rustc applies a series of Rust-specific optimizations on the MIR, ensuring efficient code generation while maintaining memory safety and preventing data races.

While these specific IRs and components are essential for enforcing memory safety and preventing data races, they also introduce significant complexity to compilation. Bugs in \rustc may weaken these guarantees and compromise Rust's memory safety.
For instance, a recent \rustc bug led to an unsound borrow check, allowing a program that should have been rejected to compile, potentially causing Use-After-Free\footnote{\url{https://github.com/rust-lang/rust/issues/132186}}. 
Despite their importance, existing tools and studies have overlooked the unique challenges posed by Rust-specific compilation mechanisms, leaving a gap in understanding their impact on testing \rustc.
To date, the only empirical study on Rust compilers is conducted by Xia et al.~\cite{xia2023understanding}. 
While it provides comprehensive statistics, it lacks an in-depth analysis. 
For example, it identifies \textit{src/test}, \textit{librustc}, and \textit{librustcdoc} are the three most error-prone modules in \rustc, yet they belong to the testing suite and standard library rather than the compiler itself. 
This misclassification may mislead our understanding of \rustc's design flaws.
Besides, there is currently limited tooling available to effectively test and improve the reliability of \rustc. 
In the open-source community, fuzzing scripts are commonly used to generate random programs for detecting crash bugs, but they often fail to identify compile-time issues like miscompilations.
In the research domain, RustSmith~\cite{RustSmith} is proposed as a program generator but provides limited support for Rust-specific features, including traits and generics. 
Other \rustc testing techniques~\cite{rustlantis,yang2024rusttwins,dewey2015fuzzing} attempt to generate MIR or use macros to generate test cases, but they can only detect a limited number of real \rustc bugs.
We consider that the lack of an effective testing tool specifically designed for \rustc stems largely from an insufficient understanding of the unique bug characteristics within \rustc.

To bridge the gap in understanding Rust-specific compiler bugs, we conduct a comprehensive quantitative and qualitative bug study on the official Rust compiler, \rustc. 
Our study is explicitly designed to focus on Rust-specific bugs, particularly those arising from the unique compilation stages that distinguish \rustc from traditional compilers.
Therefore, we selected labels related to Rust-specific IRs from Rust's official issue tracker, thus naturally excluding generic compiler bugs, such as those in the parser.
We focus on \rustc because it is the only compiler currently capable of handling large-scale Rust projects. 
Other unofficial Rust compilers, such as Rust-GCC~\cite{RustGCCg76:online}, remain in early development stages and lack the maturity for real-world use. 
Moreover, their bug histories are more related to build processes, such as cleanup~\cite{xia2023understanding}, rather than Rust-specific features.
In particular, our study answers the following research questions.

\begin{itemize}[left=0pt]
    \item \textbf{RQ1 (Bug Causes): What are the main causes of Rust-specific bugs in \rustc?} What is the frequency of these bug causes? Which stages/components in \rustc are more prone to bugs?

    \item \textbf{RQ2 (Symptoms): What are the symptoms of Rust-specific bugs in \rustc?} What is the frequency of these symptoms? What is the relationship between bug causes and symptoms?
    
    \item \textbf{RQ3 (Test Case Characteristics): What are the main characteristics of the bug-revealing test cases?} What kind of test settings are required to trigger \rustc bugs?
    
    \item \textbf{RQ4 (Status of Existing Tools): What are the existing testing techniques for \rustc?} What kind of bugs can they detect? What are their limitations? 
\end{itemize}

To answer these questions, we collect a list of issues and their corresponding pull requests from Rust's official GitHub~\cite{rustlang63:online} over the past three years.
Each bug is manually labeled with its symptoms, cause, and the compilation stage where it occurs.
Further, we identify frequently used language features and analyze their occurrences in both test cases and compilation commands.
Moreover, we investigate the existing testing tools for \rustc and discuss their advantages and limitations, as well as the challenges and difficulties in testing \rustc.

\textbf{Contributions.} 
The contributions of this paper can be summarized as follows.

\begin{itemize}[left=0pt]
    \item We manually construct a dataset of Rust-specific bugs in \rustc, covering the past three years. It includes test cases, issues, fixes, providing a foundation for our study and future research on testing and verifying \rustc.
    \item We conduct a comprehensive empirical study on Rust-specific compiler bugs to understand the diverse dimensions of \rustc bugs, including bug causes, bug-prone compilation stages, symptoms, and test case characteristics.
    \item Based on our analysis, we enumerate the implications of our findings, providing actionable suggestions for Rust users, \rustc developers, and programming language researchers to shed light on detecting \rustc bugs and improving its design.
\end{itemize}

\textbf{Summary of findings.} 
Some representative findings include: 

\begin{enumerate}[left=0pt]
\item The \rustc-specific IRs and components are prone to bugs due to the complex interplay of ownership, lifetimes, and trait resolution. In the HIR-processing component, most bugs (51.11\%) stem from type resolution and well-formedness checks, while MIR-related bugs mainly relate to MIR transformation (50.00\%).

\item Crash is the most common symptom (39.87\%), followed by correctness issues (25.91\%), where valid programs are mistakenly rejected or invalid ones are accepted. These often stem from the unique type checker and borrow checker within \rustc. While existing tools can detect many crash bugs, they struggle with deeper correctness and misoptimization bugs.

\item Key contributors to \rustc bugs include unstable features (24.25\%) and specific compilation settings or optimization levels (18.94\%). Features like trait objects often introduce edge cases that evade conventional testing, and their interactions with core language mechanisms can expose soundness and correctness issues.

\item Existing testing tools have detected only 6.07\% of non-crash bugs, likely due to gaps in program generation. Current approaches lack support for Rust-specific features like higher-order trait bounds, advanced lifetime annotations, and complex borrowing, limiting the detection of correctness-critical issues in valid Rust programs.
\end{enumerate}

\begin{table}[t]
\footnotesize
\caption{Rust-specific bug labels in \rustc and the corresponding descriptions~\cite{Labels·r6:online}.}
\label{tab:issue_labels}
\renewcommand\arraystretch{0.9}
\setlength{\tabcolsep}{1.5mm}{
\begin{tabular}{c|ccl}
\toprule
\textbf{Category}               & \textbf{Label}     & \textbf{\# Num} & \textbf{Description}  \\ \midrule
\multirow{2}{*}{HIR}            & A-HIR              & 20              & The high-level intermediate representation (HIR)                               \\
& \cellcolor[HTML]{EFEFEF}A-THIR             & \cellcolor[HTML]{EFEFEF}1               & \cellcolor[HTML]{EFEFEF}Typed HIR    \\ \hline
\multirow{6}{*}{MIR}            & A-MIR              & 43              & Mid-level IR (MIR) - \url{https://blog.rust-lang.org/2016/04/19/MIR.html}       \\
& \cellcolor[HTML]{EFEFEF}A-mir-opt          & \cellcolor[HTML]{EFEFEF}78              & \cellcolor[HTML]{EFEFEF}MIR optimizations      \\
& A-mir-opt-inlining & 23              & MIR inlining   \\
& \cellcolor[HTML]{EFEFEF}A-mir-opt-GVN      & \cellcolor[HTML]{EFEFEF}0               & \cellcolor[HTML]{EFEFEF}MIR opt Global Value Numbering (GVN)                                           \\
& A-mir-opt-nrvo     & 0               & Fixed by the Named Return Value Opt. (NRVO)                                    \\
& \cellcolor[HTML]{EFEFEF}A-stable-MIR       & \cellcolor[HTML]{EFEFEF}1               & \cellcolor[HTML]{EFEFEF}stable MIR    \\ \hline
\multirow{14}{*}{Type}           & A-type-system      & 25              & Type system    \\
& \cellcolor[HTML]{EFEFEF}A-inference        & \cellcolor[HTML]{EFEFEF}29              & \cellcolor[HTML]{EFEFEF}Type inference    \\
 & A-closures         & 29              & Closures (|…| \{ … \})   \\
 & \cellcolor[HTML]{EFEFEF}A-coercions        & \cellcolor[HTML]{EFEFEF}13              & \cellcolor[HTML]{EFEFEF}implicit and explicit \texttt{expr as Type} coercions                                 \\
 & A-const-generics   & 70              & const generics (parameters and arguments)                                      \\
 & \cellcolor[HTML]{EFEFEF}A-DSTs             & \cellcolor[HTML]{EFEFEF}0               & \cellcolor[HTML]{EFEFEF}Dynamically-sized types (DSTs)                                                 \\
  & A-zst              & 0               & Zero-sized types (ZST).    \\
& \cellcolor[HTML]{EFEFEF}A-trait-system     & \cellcolor[HTML]{EFEFEF}77              & \cellcolor[HTML]{EFEFEF}Trait system      \\
 & A-impl-trait       & 68              & Universally/existentially quantified anonymous types with static dispatch \\
 & \cellcolor[HTML]{EFEFEF}A-trait-objects    & \cellcolor[HTML]{EFEFEF}27              & \cellcolor[HTML]{EFEFEF}trait objects, vtable layout   \\
  & A-auto-traits      & 14              & auto traits (e.g., \texttt{auto trait Send \{\}})                                     \\
  & \cellcolor[HTML]{EFEFEF}A-implied-bounds   & \cellcolor[HTML]{EFEFEF}9               & \cellcolor[HTML]{EFEFEF}Implied bounds / inferred outlives-bounds                                      \\
 & A-coinduction      & 0               & Concerning coinduction, most often for auto traits                             \\
 & \cellcolor[HTML]{EFEFEF}A-coherence        & \cellcolor[HTML]{EFEFEF}14              & \cellcolor[HTML]{EFEFEF}Coherence    \\ \hline
\multirow{2}{*}{Lifetimes} & A-lifetimes        & 70              & Lifetimes / regions      \\
   & \cellcolor[HTML]{EFEFEF}A-borrow-checker   & \cellcolor[HTML]{EFEFEF}46              & \cellcolor[HTML]{EFEFEF}The borrow checker  \\ \bottomrule                                        
\end{tabular}}
\end{table}

\section{Study Methodology}\label{sec:methodology}

Our bug collection and analysis approach can be summarized in several steps.
Firstly, we perform bug data collection, which is detailed in Section~\ref{subsec:collect_bugs}. 
We collect issues and corresponding pull requests from the official Rust GitHub repository, categorized by issue labels. 
Then, in the next step of post-filtering, we filter out certain issues, such as duplicate bugs and those without test cases.
The final dataset consists of a series of bugs related to \rustc IRs within a specific time range (from 2022-01-01 to 2025-01-01), along with their corresponding test cases and fix patches.
Then, the resulting dataset serves as the foundation for our bug analysis approach, detailed in Section~\ref{subsec:analyzing_bugs}.
The bug analysis is conducted iteratively by four researchers. 
In each iteration, two researchers manually analyze a random sample of bugs, categorize them based on various aspects, and cross-validate the results.
When they can not reach a consensus, two additional researchers verify the categorization, resolving conflicts through discussion until a consensus is reached.


\subsection{Collecting Bugs and Fixes}\label{subsec:collect_bugs}

To collect bugs related to specific components of \rustc, we select issues from the official tracker with relevant labels.
The Rust team categorizes various labels in the tracker, which currently contains 814 labels in total. Labels prefixed with "A-" indicate different areas of the compiler, corresponding to various Rust language features and components, such as \textit{A-rustdoc}. 
Table~\ref{tab:issue_labels} lists the selected labels, descriptions, and occurrence counts. 
We focus on labels associated with Rust-specific IRs (HIR and MIR) and key language features, including the type system and memory management.
Since an issue can have multiple labels, the total label count exceeds the actual number of collected issues.
Although some dedicated tags are provided in the issue tracker, they are not suitable for our data collection scenario.
For example, \textit{C-bug} label includes all errors, including those related to parsing, which are beyond the scope of this study. 
Similarly, the \textit{T-compiler} label indicates issues assigned to the developer team, but also includes non-bug issues. 
In addition, issues are not necessarily tagged with these two labels simultaneously.

\begin{table}[t]
\footnotesize
\caption{Status and description of collected bugs.}
\label{tab:status}
\setlength{\tabcolsep}{0.9mm}{
\begin{tabular}{lll}
\toprule
\textbf{Status}      & \textbf{Description}        & \textbf{\# Num} \\ \midrule
Duplicate        & The bug duplicates other bugs that have already been confirmed.                                                                 & 88     \\
Not a bug        & It is not a bug because the feature is intentional and designed this way.                                                       & 28     \\
Not reproducible & When the developer confirmed the bug, it was no longer reproducible.               & 56     \\
Discussion         & (1) A question about a certain feature; (2) Suggestions for \rustc improvement, but not a bug.      & 6      \\
Exclude          & (1) Does not contain a test   case; (2) Unrelated to \rustc; (3) Not reproducible on 2021 edition. & 92     \\
\cellcolor[HTML]{EFEFEF}Valid            & \cellcolor[HTML]{EFEFEF}The bug has been confirmed as a \rustc bug, with a corresponding test case and fix.                & \cellcolor[HTML]{EFEFEF}301    \\ \hline
\textbf{Total}           & -   & \textbf{571}   \\ \bottomrule
\end{tabular}}
\end{table}

To ensure that all collected bugs have been reviewed by developers and have corresponding patches, only closed issues are selected. 
The time frame for collected issues is chosen to cover the period from January 1, 2022, to January 1, 2025. 
This period aligns with the usage time of the latest Rust 2021 Edition, which was released on October 21, 2021, while Rust 2024 is scheduled for release on February 20, 2025.
We crawled all closed issues that met the time frame and contained at least one tag listed in Table~\ref{tab:issue_labels}, resulting in a total of $571$ issues. 
Then, a preliminary manual post-filtering is performed to exclude unsuitable issues, with the classification criteria outlined in Table~\ref{tab:status}.
Specifically, The \textit{duplicate} status refers to issues that have already been reported and confirmed, typically marked by developers as "closed as a duplicate."
Issues labeled as \textit{not a bug} indicate that they are not actual bugs but rather expected behaviors.
Some issues are categorized as not \textit{reproducible} because they can no longer be reproduced, suggesting that the underlying bug has already been resolved. These issues are excluded due to the absence of an identifiable fix patch.
The \textit{discussion} category includes issues without test cases or specific symptoms and instead involves inquiries about usages or suggestions for alternative \rustc implementations.
Finally, the \textit{exclude} issues are unrelated to \rustc (e.g., documentation-related issues), or cannot be reproduced in the 2021 edition (e.g., issues that only occurred in the 2015 edition). 
Finally, a total of $301$ issues are classified as valid bugs and used for the subsequent analysis in our study.

\subsection{Analyzing Bugs}\label{subsec:analyzing_bugs}

To better understand the nature of the Rust-specific bugs in \rustc, and reduce the possibility of getting biased, we follow the existing bug analyzing approaches~\cite{chaliasos2021well,xiong2023an,Drosos2024When} and uniformly study bugs by an iterative process.
Specifically, in each iteration, we randomly select $20$ bugs, which are then independently examined by two co-authors. 
Each co-author categorizes the bugs based on their symptoms, bug causes, and compilation stages according to their individual understanding. 
Following this, they cross-validate and discuss their classifications until a consensus is reached. 
In cases where consensus cannot be achieved, the other two co-authors join the discussion to make the final decision. 
This iterative process was repeated $16$ times until all bugs were thoroughly analyzed. 
After 2-3 iterations, the independent categorization results of the first two co-authors showed a high degree of agreement, with around 95\% of the cases resulting in consistent labeling that required no further discussion.
The manual analysis process necessitated substantial domain-specific expertise in both the \rustc implementation and the Rust programming language, requiring approximately six person-months to complete.

To answer RQ1 and RQ2, the first two co-authors assign every bug to categories based on (1) its symptom, (2) the compilation stage when \rustc encounters the bug, and (3) its bug cause.
Specifically, regarding the bug symptom, we analyze the descriptions in each bug report to identify discrepancies between the expected and actual behavior of \rustc. 
For the compilation stage, we examine the fixes for each bug to pinpoint the exact compilation procedure and the component responsible for the issue. 
Regarding the bug causes, we review the test cases, fix patches, and developer discussions to determine the root causes of each bug.
To answer RQ3, we collect test cases from each issue, including Rust programs and compilation commands. We then extract their abstract syntax trees to identify frequent node types and analyze frequently used features and compilation commands.
To address RQ4, we gather all existing tools for testing \rustc and analyze the types and number of bugs they have identified. 
Additionally, we run each tool for 12 hours on a specific \rustc version to assess their effectiveness.

\section{RQ1: Bug Causes}\label{sec:rq1_causes}
In our collected bug list, each issue is linked to a corresponding fix pull request (PR). 
We analyze PR descriptions and code changes to classify the bug causes, as shown in Table~\ref{tab:bug_causes}.
Three categories are closely tied to Rust's language mechanisms: \textit{the type system}, \textit{the ownership system}, and errors from \textit{MIR optimizations}. 
Other categories include bugs from \textit{basic Rust syntax implementation in \rustc}, \textit{error handling and reporting}, and \textit{compatibility issues}.
The following sections provide a detailed discussion of each bug cause, with examples for clarification.

\begin{table}[t]
\caption{The taxonomy of bug causes.}
\label{tab:bug_causes}
\scriptsize
\setlength{\tabcolsep}{0.4mm}{
\begin{tabular}{c|c|l|c|c}
\toprule
\textbf{Category}                        & \textbf{Subcategory}   & \textbf{Description}      & \textbf{\# Bugs} & \textbf{Ratio}   \\ \hline
\multirow{9}{*}{\begin{tabular}[c]{@{}c@{}}Type \\ System\\  Errors\end{tabular}}      & Trait \& Bound         & \parbox{9cm}{\vspace{0.05cm}The errors were caused by \rustc's handling of traits and its enforcement of type parameter constraints, such as requiring specific traits or conditions.\vspace{0.05cm}}                               & 37               & 12.29\%          \\ 
 & \cellcolor[HTML]{EFEFEF}Opaque types         & \cellcolor[HTML]{EFEFEF}\parbox{9cm}{\vspace{0.05cm}The errors were caused by issues within \rustc's handling of opaque types, which rely on the ownership system, zero-cost abstractions, and the design of generics and traits.\vspace{0.05cm}}        & \cellcolor[HTML]{EFEFEF}38               & \cellcolor[HTML]{EFEFEF}12.62\%          \\ 
 & New solver           & \parbox{9cm}{\vspace{0.05cm}The errors are caused due to the interaction between \rustc's new solver, which is designed to improve trait-bound resolution and reduce workload, and the existing old solver.\vspace{0.05cm}} & 7               & 2.33\%           \\ 
 & \cellcolor[HTML]{EFEFEF}Well-formedness & \cellcolor[HTML]{EFEFEF}\parbox{9cm}{\vspace{0.05cm}The errors were caused by \rustc's well-formedness checking, including ownership, lifetime, type system, and the borrow checker.\vspace{0.05cm}}     & \cellcolor[HTML]{EFEFEF}9                & \cellcolor[HTML]{EFEFEF}2.99\%           \\ 
 & \textbf{Subtotal}      & \textbf{-}             & \textbf{91}      & \textbf{30.23\%} \\ \hline
\multirow{4}{*}{\begin{tabular}[c]{@{}c@{}}Ownership   \\ \& Lifetime \\ Errors\end{tabular}} & Borrow \& Move         & \parbox{9cm}{\vspace{0.05cm}The errors were caused by issues in implementing the ownership model, which ensures memory safety and concurrency safety through the move and borrow semantics.\vspace{0.05cm}}                    & 7                & 2.33\%           \\ 
  & \cellcolor[HTML]{EFEFEF}Lifetime               & \cellcolor[HTML]{EFEFEF}\parbox{9cm}{\vspace{0.05cm}The errors were caused by issues in \rustc's lifetime checking, which ensures that every reference is valid and does not outlive the data it points to.\vspace{0.05cm}}    & \cellcolor[HTML]{EFEFEF}34               & \cellcolor[HTML]{EFEFEF}11.30\%          \\ 
  & \textbf{Subtotal}      & \textbf{-}            & \textbf{41}      & \textbf{13.62\%} \\ \hline
\multirow{3}{*}{\begin{tabular}[c]{@{}c@{}}MIR \\ Optimization \\ Errors\end{tabular}}      & \cellcolor[HTML]{EFEFEF}\begin{tabular}[c]{@{}c@{}}Wrong \\ implementations\end{tabular}  & \cellcolor[HTML]{EFEFEF}\parbox{9cm}{\vspace{0.05cm}
The errors were caused by incorrect implementations of \rustc's MIR-based optimizations (e.g., constant propagation, dead code elimination, inlining).\vspace{0.05cm}}              & \cellcolor[HTML]{EFEFEF}34               & \cellcolor[HTML]{EFEFEF}11.30\%   \\ 
 & Missing cases        & \parbox{9cm}{\vspace{0.05cm}Some specific corner cases of the optimization algorithm were not considered thoroughly.\vspace{0.05cm}}                      & 12               & 3.99\%           \\ 
 & \cellcolor[HTML]{EFEFEF}\textbf{Subtotal}      & \cellcolor[HTML]{EFEFEF}\textbf{-}             & \cellcolor[HTML]{EFEFEF}\textbf{46}      & \cellcolor[HTML]{EFEFEF}\textbf{15.28\%} \\ \hline
 
\multirow{6}{*}{\begin{tabular}[c]{@{}c@{}}General \\ Errors\end{tabular}} & Basic structure                &       \parbox{9cm}{\vspace{0.05cm}Bugs caused by \rustc errors in processing features like closures and internal data structures.\vspace{0.05cm}}                          & 38            & 12.62\%        \\ 
& \cellcolor[HTML]{EFEFEF}\begin{tabular}[c]{@{}c@{}}Error handling \\ \& Reporting  \end{tabular}              & \cellcolor[HTML]{EFEFEF}\parbox{9cm}{\vspace{0.05cm}The errors were caused by \rustc's failure to handle exceptional cases properly or its misprocessing of reports, leading to misleading error messages or incorrect error locations.\vspace{0.05cm}}      & \cellcolor[HTML]{EFEFEF}75              & \cellcolor[HTML]{EFEFEF}24.92\%      \\ 
& Compatibility        & \parbox{9cm}{\vspace{0.05cm}The bugs were triggered by certain operating systems, bugs in the back-end LLVM, or errors specific to the Rust edition.\vspace{0.05cm}}    & 10            & 3.32\% \\ 
& \cellcolor[HTML]{EFEFEF}\textbf{Subtotal}      & \cellcolor[HTML]{EFEFEF}\textbf{-}             & \cellcolor[HTML]{EFEFEF}\textbf{123}      & \cellcolor[HTML]{EFEFEF}\textbf{40.86\%} \\ \bottomrule

\end{tabular}}
\end{table}

\subsection{Type System Errors} \label{subsec:type_system_errors}
Type system errors are a major cause of bugs in \rustc, accounting for 30.23\% of all cases. 
These issues stem from \rustc handling Rust's complex type mechanism, which emphasizes zero-cost abstractions, allowing high-level, expressive code without runtime overhead. 
For example, traits enable polymorphism, and generics allow code to operate on multiple types while maintaining type safety.
However, their interaction with Rust's other mechanism such as the ownership model introduces significant complexity, often leading to intricate type relationships and related bugs.
We classify an error as a type system error if \rustc fails to correctly handle Rust's type mechanisms, leading to incorrect behavior or a compilation failure.
Type system-related bugs belong to one of the following groups: \textit{(1)trait \& bound related errors}, \textit{(2) opaque types related errors}, \textit{(3) new solver related errors}, or \textit{(4) well-formedness related errors}.

\textbf{\textit{Trait \& Bound Related Errors:}}
Trait-related errors account for 12.29\% of all bugs. Traits define shared behaviors across types, while bounds constrain the types that can be used with generics. These bounds work with traits to ensure type safety and enable polymorphism.  
Errors in this category occur when \rustc struggles to resolve trait bounds or apply constraints during type inference or checking. Typically, this happens when \rustc fails to match types to their associated trait bounds, leading to incorrect type assignments or failure to resolve the required traits.

\textbf{\textit{Opaque types Related Errors:}}
Opaque types allow defining a type alias that only exposes certain traits as its interface. The actual concrete type is inferred from its usage in the code context~\cite{OpaqueTy48:online}. 
Examples include types introduced by \colortexttt{mybrown}{impl Trait} and associated types within traits. For \rustc, handling opaque types requires resolving these types and their associated properties during type checking and inference while maintaining their abstraction across different scopes. 
Errors in this category, which account for 12.62\% of all causes, occur when \rustc encounters difficulties in properly resolving opaque types or their associated properties, often due to scope-related issues. 
These challenges can lead to incorrect behavior, such as type mismatches or compilation failures, revealing flaws in \rustc's type resolution for opaque types.

\textbf{\textit{New solver Related Errors:}}
The Rust team has been actively developing and integrating a new trait solver to replace some of the existing core implementations~\cite{Nextgent9:online}. 
This effort aims to address unsoundness issues in the previous solver and enhance compilation efficiency.
Currently, both the old and new trait solvers coexist within \rustc, leading to challenges during the transition.
Errors in this category, accounting for 2.33\% of all causes, usually result from issues in the new trait solver, especially when resolving complex trait bounds.

\textbf{\textit{Well-formedness Related Errors:}}
Well-formedness (WF)~\cite{Wellform75:online} ensures that declarations in a Rust program follow its language's rules, validating types, bounds, and relationships. 
The WF checker generates a logical goal for each declaration and attempts to prove it using the type system's rules. If successful, the declaration is deemed well-formed; otherwise, an error is reported.
Errors in this category, accounting for 2.99\% of all causes, arise from \rustc improperly processing WF checking, leading to incorrect behaviors or ICEs when validating the well-formedness.

\begin{figure}[htbp]
  \centering
  \includegraphics[width = \linewidth]{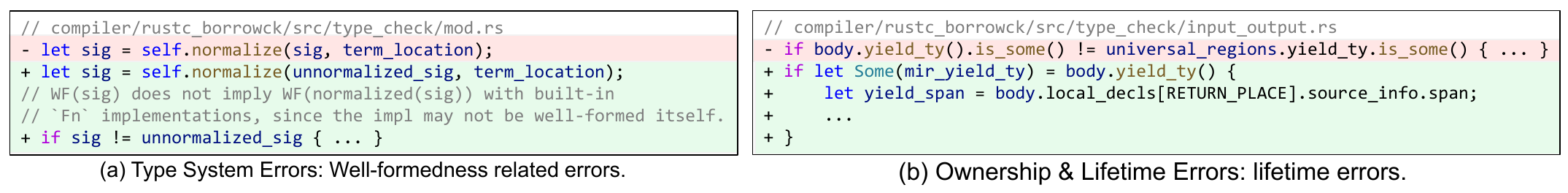}
  \vspace{-10pt}
  \caption{Two snippets of fix patch for explaining ownership \& lifetime bug cause (\href{https://github.com/rust-lang/rust/pull/118882}{\textcolor{Violet}{PR 118882}} and \href{https://github.com/rust-lang/rust/pull/119563}{\textcolor{Violet}{PR 119563}}).}
  \vspace{-10pt}
  \label{fig:cause_type_ownership}
\end{figure}

\textbf{Example.}
The patch in Figure~\ref{fig:cause_type_ownership} (a) addresses a WF-related error (tracked as \href{https://github.com/rust-lang/rust/issues/118876}{\textcolor{Violet}{Issue 118876}}) caused by incorrect WF checking for built-in traits.
The built-in \colortexttt{mybrown}{Fn*} traits, including \colortexttt{mybrown}{Fn}, \colortexttt{mybrown}{FnMut}, and \colortexttt{mybrown}{FnOnce}, allow closures to be used like function pointers, passed as arguments, or stored in structs.
Before explaining the bug cause, we first clarify some definitions. 
The unnormalized signature refers to function signatures that may include unresolved associated types, whereas the normalized signature resolves all associated types to their concrete definitions. 
Rust's type system assumes that if a type is well-formed, its normalized form is also well-formed. 
As a result, \rustc only checks the WF of the unnormalized signature and ignores the normalized form during type checking.
However, this assumption is violated because the implementations of built-in \colortexttt{mybrown}{Fn*} traits do not explicitly declare certain required lifetime bounds, particularly the \colortexttt{mybrown}{\textquotesingle s: \textquotesingle static} bound. Consequently, \rustc fails to enforce these implicit lifetime bounds, leading to an unexpected compiler behavior.
The patch in Figure~\ref{fig:cause_type_ownership} (a) adds checks for the normalized signature to ensure that all associated types are resolved and necessary lifetime bounds are explicitly declared. 
This ensures that \rustc applies the same WF rules to both built-in \colortexttt{mybrown}{Fn*} traits and user-defined traits.

\subsection{Ownership \& Lifetime Errors} \label{subsec:ownership_errors}
Rust's ownership and lifetime system is a core language feature, ensuring memory safety without needing a garbage collector. 
Bugs in \rustc related to ownership and lifetime errors account for 13.62\% of all causes.
Specifically, \rustc checks whether references are valid for their expected lifetimes, ensuring that mutable and immutable references do not conflict, and enforcing the ownership rules that prevent data from being accessed after it has been moved or dropped.
We classify an error as an ownership and lifetime error when Rust's ownership model fails, causing compilation issues. 
These bugs belong to one of the following groups: (1) \textit{borrow \& move related errors} or (2) \textit{lifetime related errors}.

\textbf{\textit{Borrow \& Move Related Errors:}}
The borrow and move mechanisms are fundamental to Rust’s ownership system, yet bugs arising from them are relatively rare, accounting for only 2.33\% of all identified causes.
The borrow model enables references to a value without transferring ownership, permitting either multiple immutable references or a single mutable reference, but never both simultaneously.
The move model, in contrast, transfers ownership of a value, rendering the original variable invalid and preventing further use.
Bugs in this category typically stem from \rustc mismanaging mutable and immutable borrowing or incorrectly tracking ownership transfers.

\textbf{\textit{Lifetime Related Errors:}}
The lifetime is a key feature of Rust's ownership system, which describes the scope for which a reference is valid, preventing issues like dangling references or data races.
The bugs caused by lifetime-related errors account for 11.30\% of all.
The borrow checker in \rustc utilizes lifetimes to track the validity of references and enforce that they do not outlive the referenced data.
The errors caused by this category are typically because \rustc improperly infer or check the lifetimes of references.

\textbf{Example.}
The patch in Figure~\ref{fig:cause_type_ownership} (b) shows an example of lifetime-related errors.
In the test case (tracked as \href{https://github.com/rust-lang/rust/issues/119564}{\textcolor{Violet}{Issue 119564}}), \colortexttt{mybrown}{coroutines} for asynchronous programming are utilized.
Unlike traditional functions, \colortexttt{mybrown}{coroutines} in Rust allow execution to be paused and resumed at different points, forming an implicit state machine. 
This mechanism introduces challenges for the borrow checker, as it must ensure that all references captured inside the \colortexttt{mybrown}{coroutine} remain valid across suspension points.
However, in this case, \rustc failed to properly enforce lifetime constraints on values produced by \colortexttt{mybrown}{yield}, allowing a yielded value to be assigned a stricter lifetime than it should have. 
Since \colortexttt{mybrown}{yield} effectively acts as a suspension point, any borrowed reference tied to it must remain valid when the \colortexttt{mybrown}{coroutine} resumes. Without proper checks, this could lead to dangling references or memory safety violations.
The patch shown in Figure~\ref{fig:cause_type_ownership} (b) improves soundness in \rustc's coroutine handling by enforcing stricter lifetime checks at yield and resumption points. 
When a yield expression is detected, \rustc captures the \texttt{yield\_span} to determine the scope of the yielded value. 
Then, \rustc uses this span to perform further checks, ensuring that \colortexttt{mybrown}{coroutines} correctly enforce lifetime constraints.

\subsection{MIR Optimization Errors}
MIR optimization in \rustc refines MIR to enhance performance and reduce resource consumption.
These optimizations, including constant folding, dead code elimination, and loop unrolling, refine the code before it is passed to the backend compiler. 
While most algorithms have been implemented within classic compilers, applying them to MIR can introduce subtle interactions and edge cases.
Bugs arising from these challenges, categorized as MIR optimization errors, account for 15.28\% of all causes.
An MIR optimization error occurs when incorrect transformation or optimization causes misbehavior or compilation failure.
These bugs fall into two categories: \textbf{(1) \textit{wrong implementations}}, where \rustc incorrectly implement the intended transformations (11.3\%), and \textbf{(2) \textit{missing cases}}, where certain corner cases or program patterns are not properly addressed, leading to incomplete optimizations (3.99\%). 
From our study, most MIR optimization bugs require modifications to the algorithm's logic, rather than merely fixing a minor overlooked case.

\begin{figure}[htbp]
  \centering
  \includegraphics[width = \linewidth]{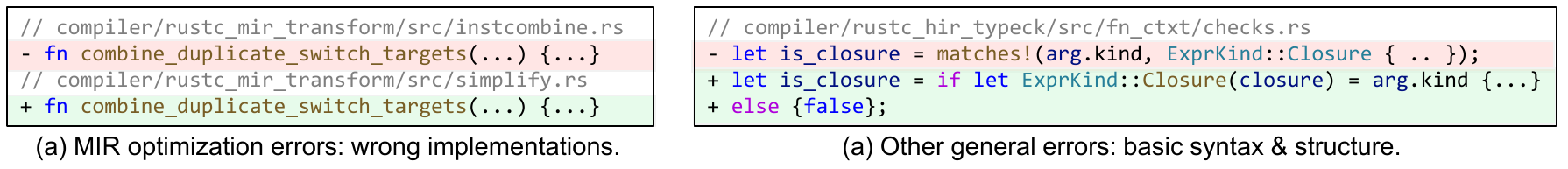}
  \caption{The fix patch for explaining MIR optimization bug cause \href{https://github.com/rust-lang/rust/pull/110569}{\textcolor{Violet}{PR 110569}}) and general errors \href{https://github.com/rust-lang/rust/pull/112266}{\textcolor{Violet}{PR 112266}}).}
  \label{fig:cause_mir_general}
\end{figure}

\textbf{Example.}
Figure~\ref{fig:cause_mir_general} (a) illustrates an example of incorrect MIR optimization. The bug (tracked as \href{https://github.com/rust-lang/rust/issues/110551}{\textcolor{Violet}{Issue 111355}}) occurs when inlining results in redundant unreachable blocks.
It is caused by the interaction between two key MIR optimization passes: \colortexttt{mybrown}{InstCombine}, which simplifies instructions by combining constant expressions and redundant operations, and \colortexttt{mybrown}{SimplifyCfg}, which simplifies control flow graphs by removing unnecessary branches and loops.
Initially, the function responsible for merging duplicate targets was placed within \colortexttt{mybrown}{InstCombine}, but this placement was ineffective because \colortexttt{mybrown}{InstCombine} runs before \colortexttt{mybrown}{SimplifyCfg}. Since duplicate unreachable blocks are only introduced after \colortexttt{mybrown}{SimplifyCfg} is applied, the function was executed too early to have the intended effect.
The patch corrects this by relocating the function, ensuring it properly merges duplicate unreachable blocks when they actually appear.

\subsection{General Errors}
The remaining bug causes are not directly tied to core language features but instead, result from more fundamental issues in how \rustc processes certain constructs, handles edge cases, or interacts with its backend systems. 
These errors can stem from various issues within \rustc's internal logic, structure, or interaction with external components. These bugs account for 40.86\% of all causes.
We classify an error as a general error when flaws in \rustc's design or implementation cause expected compilation behaviors.
General errors in \rustc can be classified into three categories: \textbf{(1) \textit{basic structure errors}}, where \rustc incorrectly processes fundamental constructs, such as closures or internal data structures (12.62\%); \textbf{(2) \textit{error handling and reporting issues}}, where exceptional cases or error reports are mishandled, leading to misleading messages or incorrect error locations (24.92\%); and \textbf{(3) \textit{compatibility issues}}, where bugs arise from specific operating system configurations, backend LLVM problems, or Rust edition-specific errors (3.32\%).

\textbf{Example.}
Figure~\ref{fig:cause_mir_general} (b) illustrates an example of basic structure errors, which is a regression in Rust 1.70 (tracked as \href{https://github.com/rust-lang/rust/issues/112225}{\textcolor{Violet}{Issue 112225}}) affecting type inference in argument-position closures and \colortexttt{mybrown}{async} blocks. 
The issue arises from how \rustc evaluates \colortexttt{mybrown}{async} blocks, where improper closure handling leads to incorrect type resolution.
Unlike regular functions, \colortexttt{mybrown}{async} blocks are implicitly transformed into state machines, which affects closure inference and evaluation order. 
This transformation caused \rustc to misidentify closures in arguments, leading to inference failures.
The patch adds an explicit check to verify whether an argument is a closure, preventing misclassification.

\subsection{Bug Prone Compilation Stages}\label{subsec:stages}

\begin{figure}[t]
    \centering
    \subfigure[Bug distribution across \rustc compilation stages.]{
        \label{subfig:stages_distribution}
        \includegraphics[width=0.44\linewidth]{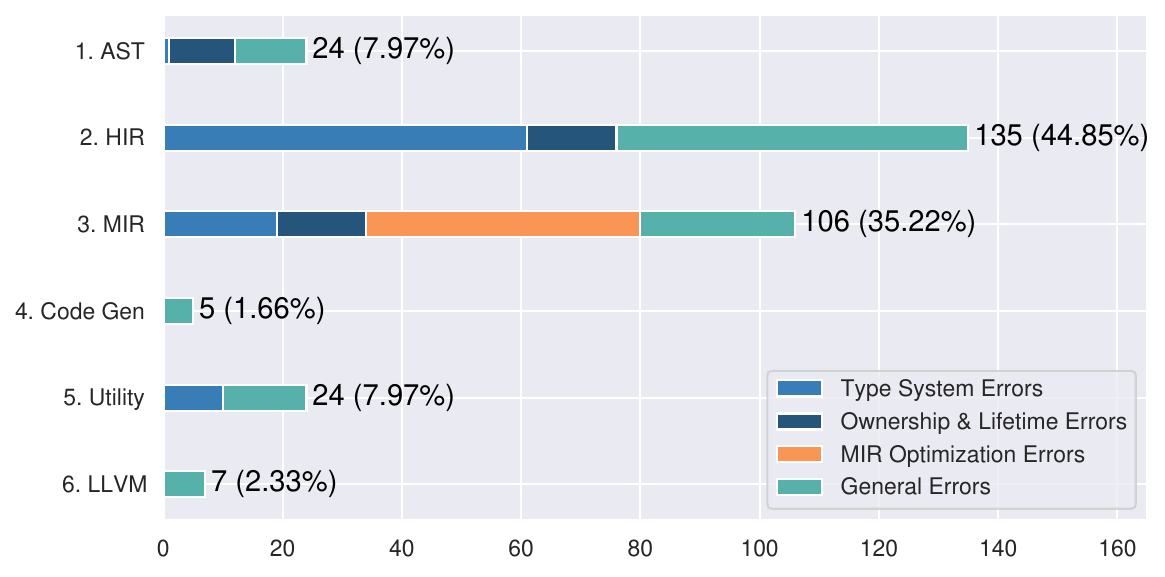}
    }
    \subfigure[Bug distribution across HIR and MIR components.]{
        \label{subfig:hir_mir_distribution}
        \includegraphics[width=0.52\linewidth]{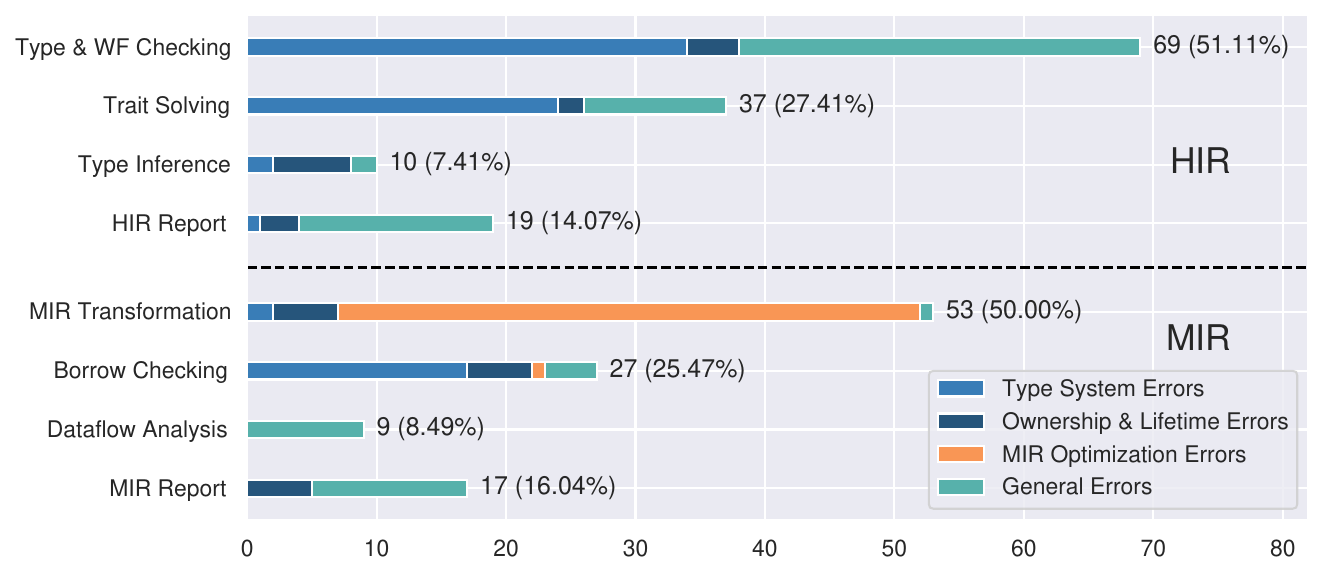}
    }
    \caption{Comprehensive analysis of bug distribution in \rustc pipeline and its HIR/MIR components.}
    \label{fig:stage_cause}
\end{figure}

The workflow of \rustc involves several specific components, including HIR and MIR, as well as various specialized checks and analyses based on these IRs that support Rust's unique memory management system.
To investigate the stages of \rustc compiler pipeline prone to bugs, we decompose its workflow and divide it into several core stages. We then quantify the error rates at each stage and analyze the underlying causes.
In some cases, a bug involves modifications across multiple stages. To handle such cases, we identify all affected modules in the fixing PR and trace the root cause to the stage where the error originates.
Figure~\ref{subfig:stages_distribution} provides an overview of the distribution of bug causes across different compilation stages. 
General errors appear throughout all stages, while MIR optimization bugs predominantly occur in the MIR-processing stage. 
Beyond HIR-processing and MIR-processing, most bugs stem from general errors.

To further understand the bugs triggered in the core HIR and MIR components, we subdivide them and investigate their bug causes, as shown in Figure~\ref{subfig:hir_mir_distribution}. 
A closer look at the HIR and MIR processing stages reveals that bugs related to the type system and ownership mechanisms are spread across multiple components rather than being isolated to a single stage.
Regarding Figure~\ref{subfig:hir_mir_distribution}, most components contain bugs caused by type system errors. 
For instance, issues in \textit{type \& WF checking} may allow invalid types, while errors in \textit{trait solving} can lead to unexpected type mismatches.
Bugs in \textit{MIR transformation} and \textit{borrow checking} can also stem from type system errors. 
This is partly because some WF checks are performed during borrow checking, as certain lifetime information may still be incomplete during the HIR-based type-checking phase.
Similarly, multiple components are affected by ownership and lifetime errors.
In the \textit{type \& WF checking} and \textit{type inference} components, incorrectly inferred types and constraints can lead to unsound borrowing rules. 
Additionally, incorrect trait resolution in the \textit{trait solving} may introduce errors that propagate to later stages, ultimately affecting ownership analysis.
Errors in \textit{borrow checking} can directly cause ownership-related issues. 
Furthermore, the \textit{MIR transformation} involves optimization algorithms related to lifetimes, which can also introduce related issues.
Regarding \textit{error reporting} in both the HIR and MIR components, most bugs stem from general programming errors, especially improper error handling, which can misclassify the root causes of compilation failures.

\section{RQ2: Bug Symptoms}\label{sec:rq2_symptoms}


To categorize the bug symptoms of \rustc, we manually review bug descriptions from GitHub's bug reports and analyze the discrepancies between expected and actual behaviors.
Specifically, we categorized the bugs into five distinct bug symptom categories: \textit{Crash}, \textit{Correctness Issues}, \textit{Miscompilation}, \textit{Diagnostic Issues} and \textit{Misoptimization}. 
The distribution of bug symptoms is shown in Table~\ref{tab:symptoms}.
Among them, crashes are the most prevalent, accounting for 39.87\% of cases, followed by correctness issues (25.91\%) and diagnostic issues (19.27\%). Miscompilation and misoptimization are less common, making up 9.97\% and 4.98\%, respectively. 

\begin{table}[htbp]
\footnotesize
\caption{Distribution of bug symptoms and the distribution of bug symptoms per cause.}
\label{tab:symptoms}
\setlength{\tabcolsep}{0.65mm}{
\begin{tabular}{c|c|c|c||cccc}
\toprule
\multicolumn{2}{c|}{\textbf{Symptoms}}        & \multicolumn{2}{c||}{\textbf{Occurrence}} & \begin{tabular}[c]{@{}c@{}}\textbf{Type}\\ \textbf{System}\end{tabular} & \begin{tabular}[c]{@{}c@{}}\textbf{Ownership}\\ \textbf{\& Lifetime}\end{tabular} & \begin{tabular}[c]{@{}c@{}}\textbf{MIR}\\ \textbf{Optimization}\end{tabular} & \begin{tabular}[c]{@{}c@{}}\textbf{General}\\ \textbf{Errors}\end{tabular} \\ \hline
\multirow{3}{*}{\textbf{Crash}}              & Front-end Panic (valid)     & 42 (13.95\%) & \multirow{3}{*}{\begin{tabular}[c]{@{}c@{}}120 \\ (39.87\%)\end{tabular}} &  \multirow{3}{*}{\begin{tabular}[c]{@{}c@{}} 30\\ (25.00\%) \end{tabular}} & \multirow{3}{*}{\begin{tabular}[c]{@{}c@{}}3\\ (2.50\%)\end{tabular}} & \multirow{3}{*}{\begin{tabular}[c]{@{}c@{}}19\\ (15.83\%)\end{tabular}}  & \multirow{3}{*}{\begin{tabular}[c]{@{}c@{}}\cellcolor[HTML]{EFEFEF}\textbf{68} \\ \cellcolor[HTML]{EFEFEF}\textbf{(56.67\%)}\end{tabular}} \\
  & Front-end Panic (invalid) & 75 (24.92\%) &                          \\
  & Back-end Crash                      & 3 (1.00\%)  &  &  & &  & \\ \hline
\multirow{2}{*}{\textbf{Correctness Issues}} & Completeness Issues                 & 56 (18.60\%) & \multirow{2}{*}{\begin{tabular}[c]{@{}c@{}} 78 \\ (25.91\%)\end{tabular}} &  \multirow{2}{*}{\begin{tabular}[c]{@{}c@{}}\cellcolor[HTML]{EFEFEF}\textbf{43} \\ \cellcolor[HTML]{EFEFEF}\textbf{(55.13\%)}\end{tabular}} & \multirow{2}{*}{\begin{tabular}[c]{@{}c@{}}17 \\ (21.79\%)\end{tabular}} & \multirow{2}{*}{\begin{tabular}[c]{@{}c@{}}7 \\ (8.97\%)\end{tabular}} & \multirow{2}{*}{\begin{tabular}[c]{@{}c@{}}11 \\ (14.10\%)\end{tabular}} \\
    & Soundness Issues                    & 22 (7.31\%)  &  &  & &  &      \\  \hline
\multirow{2}{*}{\textbf{Miscompilation}}     & Inconsistent Output Issues          & 18 (5.98\%)  & \multirow{2}{*}{\begin{tabular}[c]{@{}c@{}} 30 \\(9.97\%)\end{tabular}} & \multirow{2}{*}{\begin{tabular}[c]{@{}c@{}}5 \\ (16.67\%)\end{tabular}} & \multirow{2}{*}{\begin{tabular}[c]{@{}c@{}}4\\ (13.33\%)\end{tabular}} & \multirow{2}{*}{\begin{tabular}[c]{@{}c@{}}\cellcolor[HTML]{EFEFEF}\textbf{12}\\ \cellcolor[HTML]{EFEFEF}\textbf{(40.00\%)}\end{tabular}} & \multirow{2}{*}{\begin{tabular}[c]{@{}c@{}}9 \\ (30.00\%)\end{tabular}} \\ 
  & Safe Rust Causes UB         & 12 (3.99\%)  &  &&&&   \\ \hline
\multirow{2}{*}{\textbf{Diagnostic Issues}}  & Incorrect Warning/Error             & 20 (6.64\%)  & \multirow{2}{*}{\begin{tabular}[c]{@{}c@{}}58 \\ (19.27\%)\end{tabular}} & \multirow{2}{*}{\begin{tabular}[c]{@{}c@{}}12\\ (20.69\%)\end{tabular}} & \multirow{2}{*}{\begin{tabular}[c]{@{}c@{}}16\\(27.59\%)\end{tabular}} & \multirow{2}{*}{\begin{tabular}[c]{@{}c@{}}1\\(1.72\%)\end{tabular}} & \cellcolor[HTML]{EFEFEF}\multirow{2}{*}{\begin{tabular}[c]{@{}c@{}}\textbf{29}\\ \cellcolor[HTML]{EFEFEF}\textbf{(50.00\%)}\end{tabular}}\\
  & Improper Fix Suggestion          & 38 (12.62\%) & &  & &  &    \\ \hline
\multicolumn{2}{c|}{\textbf{Misoptimization}}  & 15 (4.98\%)    & \begin{tabular}[c]{@{}c@{}}15 \\ (4.98\%) \end{tabular}  & \begin{tabular}[c]{@{}c@{}}1 \\ (6.67\%)\end{tabular} & \begin{tabular}[c]{@{}c@{}}1 \\ (6.67\%)\end{tabular} &\cellcolor[HTML]{EFEFEF}\begin{tabular}[c]{@{}c@{}}\textbf{7} \\ \textbf{(46.67\%)}\end{tabular} & \begin{tabular}[c]{@{}c@{}}6 \\ (40.00\%)\end{tabular}  
\\ \bottomrule 
\end{tabular}}
\end{table}

\subsection{Crash}
Similar to all software systems, \rustc also suffers from crashes.
Among the bugs we collected, 36.51\% involve crash errors.
Based on the compilation stage where the crash occurs, we categorize them into front-end panics and back-end crashes. 

\textbf{\textit{Front-end Panic:}}
In Rust, a panic occurs on an unrecoverable error, followed by a cleanup operation before termination.
In \rustc, an internal compiler error (ICE) often manifests as a panic, indicating that \rustc has encountered an unexpected state or an unhandled scenario.
The front-end panic accounts for 38.87\% of all observed symptoms. Among them, 13.95\% are triggered by valid programs, and 24.92\% are triggered by invalid programs.

\textbf{\textit{Back-end Crash:}}
Back-end crashes happen due to low-level failures like segmentation faults (SIGSEGV) or abnormal terminations (SIGABRT), often linked to issues with code generation or LLVM. 
The proportion of back-end crashes is notably minimal, accounting for 1\% of all cases.

\textbf{Bug cause analysis.}
The primary causes of crash bugs in \rustc are general errors (56.67\%), such as inadequate error handling and compatibility issues. When \rustc encounters an unexpected program state, its error recovery mechanisms may be incomplete, leading to an ICE instead of graceful handling.
The second major category involves the type system (25.00\%) and MIR optimization (15.83\%). The complexity of type checking and trait resolution can introduce subtle inconsistencies, especially with advanced generics and associated types. Additionally, since MIR serves as the bridge between high-level Rust code and low-level machine code, incorrect optimizations or misinterpretations of type transformations at this stage can also lead to ICE.
Finally, ownership and lifetime errors account for 2.5\% of crash bugs. As most violations in this area are caught at compile time, incorrect checks are more likely to cause correctness issues rather than immediate crashes.

\begin{figure}[htbp]
  \centering
  \includegraphics[width = \linewidth]{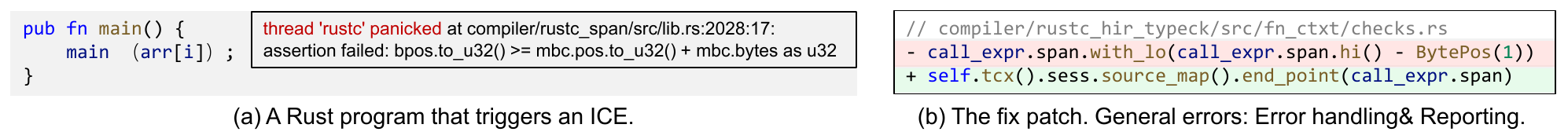}
  \caption{The example of a crash bug \href{https://github.com/rust-lang/rust/issues/128717}{\textcolor{Violet}{(Issue 128717)}} and corresponding fix patch \href{https://github.com/rust-lang/rust/pull/128864}{\textcolor{Violet}{(PR 128864)}}.}
  \label{fig:exp_ice}
\end{figure}

\textbf{Example.}
Figure~\ref{fig:exp_ice} (a) shows a code snippet that triggers an ICE, which is caused by a \textit{general error}.
This code incorrectly passes an argument into the main function and uses a multi-byte brace as the closing delimiter. 
When \rustc detects that the main function involves parameters, it attempts to provide a fix suggestion and shifts one byte to remove the extra parameter.
However, \rustc fails to handle multi-byte characters because it assumes that every closing delimiter is a single byte. 
This misalignment violates Unicode boundaries, triggering an assertion failure.
Figure~\ref{fig:exp_ice} (b) presents the fix patch, which corrects the positioning approach by eliminating the use of \texttt{BytePos}.

\subsection{Correctness Issues}
Correctness bugs occur when \rustc fails to enforce Rust's syntax rules, leading to the unintended rejection or acceptance of programs, undermining its ability to accurately validate Rust programs.
The correctness issues account for 25.91\% of all cases. Among them, 18.60\% are triggered by completeness issues, and 7.31\% are triggered by soundness issues.

\textbf{\textit{Completeness Issues:}}
Completeness bugs occur when \rustc fails to compile a syntactically and semantically valid Rust program as defined by the language specification. These bugs typically manifest when \rustc incorrectly rejects such a program, either by displaying a false error message or failing to complete compilation.

\textbf{\textit{Soundness Issues:}}
Soundness bugs refer to situations where \rustc mistakenly accepts programs that should be rejected due to violating language rules.
Rust is known for its strict rules around syntax and semantics, and soundness bugs occur when \rustc incorrectly allows code that violates these rules to compile successfully. 

\textbf{Bug cause analysis.}
Correctness bugs in \rustc primarily stem from issues in the type system (55.13\%) and ownership management (21.79\%). 
Unlike crash bugs, correctness issues are not as immediately apparent, as they often result from logical flaws in \rustc's core checking mechanisms rather than explicit failures. 
Other causes, such as MIR optimization errors (8.97\%) and general errors (14.10\%), are relatively less common. MIR optimization bugs can introduce subtle miscompilations when incorrect transformations alter program semantics, particularly in aggressive optimization scenarios. General errors, including missed edge cases in \rustc logic, may propagate inconsistencies, leading to undetected violations of Rust's safety guarantees.

\begin{figure}[htbp]
  \centering
  \includegraphics[width = \linewidth]{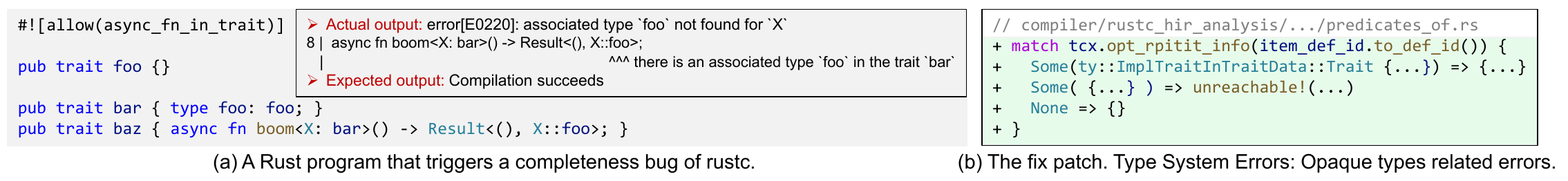}
  \caption{The example of a correctness issue \href{https://github.com/rust-lang/rust/issues/132372}{\textcolor{Violet}{(Issue 132372)}} and corresponding fix patch \href{https://github.com/rust-lang/rust/pull/132373}{\textcolor{Violet}{(PR 132373)}}.}
  \label{fig:exp1}
\end{figure}

\textbf{Example.}
Figure~\ref{fig:exp1} (a) presents a test case that exposes a \textit{correctness issue}, which is caused by handling \textit{opaque types} and Return-Position Impl Trait in Trait (\colortexttt{mybrown}{RPITITs}), categorized under \textit{type system errors}. 
The test case defines three public traits: \texttt{foo}, \texttt{bar}, and \texttt{baz}. The \texttt{baz} trait includes an asynchronous method, \texttt{boom}, which is generic over a type \texttt{X} constrained to implement \texttt{bar}. 
Here, the asynchronous method can be defined in the trait because the corresponding unstable feature is enabled.
While this code previously compiled successfully, it now fails with the latest \rustc version.
Although the test case does not explicitly use \colortexttt{mybrown}{impl Trait}, the \colortexttt{mybrown}{async} function implicitly returns \colortexttt{mybrown}{impl Future<Output=T>}, thereby involving RPITITs. 
The root cause of the bug is that RPITITs are incorrectly assigned the \textit{def id} of a Generalized Associated Type (GAT) instead of the correct opaque type identifier.
Consequently, shorthand projections such as \colortexttt{mybrown}{T::Assoc} fail to resolve properly.
The patch in Figure~\ref{fig:exp1} (b) corrects this by modifying \rustc to detect cases where an item originates from \colortexttt{mybrown}{RPITIT} lowering and ensuring that queries are forwarded to the appropriate item. 
As this case demonstrates, flaws in the complex type system can lead to correctness issues, underscoring the challenges of maintaining a reliable type system in Rust.

\subsection{Miscompilation}
Miscompilation bugs occur when \rustc generates incorrect machine code or behaves unexpectedly during compilation, leading to incorrect program execution. 
Miscompilation issues are particularly important, as they may compromise the safety and performance guarantees that Rust provides to its users.
Bugs classified as miscompilation account for 9.97\% of the total.

\textbf{\textit{Inconsistent Output Issues:}}
These bugs arise when \rustc produces different outputs based on compilation levels or optimization settings. Rust's debug and release modes apply varying optimizations, but miscompilation can cause inconsistencies in both the generated machine code and the program's execution results across configurations.
Bugs classified as inconsistent output issues account for 5.98\% of all symptoms.

\textbf{\textit{Safe Rust Program Causes Undefined Behaviors:}}
This symptom is unique to \rustc, as it guarantees that safe Rust code does not cause undefined behavior (UB). UB occurs when a program executes code that is not defined by the language specification, leading to unpredictable or incorrect results~\cite{Undefine65:online}.
However, a \rustc bug is triggered if a safe Rust program causes UB, which accounts for 3.99\% of all symptoms.

\textbf{Bug cause analysis.}
Miscompilation bugs in \rustc are primarily caused by MIR optimization errors (40.00\%). Faulty optimization logic can lead to semantic differences between optimized and unoptimized code, directly affecting program correctness. 
General errors account for 30.00\%, as mistakes in internal data structures or improper handling of basic syntax can propagate through the compilation process, leading to incorrect code generation.  
Other causes, including type system issues (16.67\%) and ownership-related errors (13.33\%), are relatively less common. 
Type system bugs may lead to miscompilations due to incorrect type inference or trait resolution. Similarly, ownership errors could result in unintended memory access patterns, potentially causing miscompilations.

\begin{figure}[htbp]
  \centering
  \includegraphics[width = \linewidth]{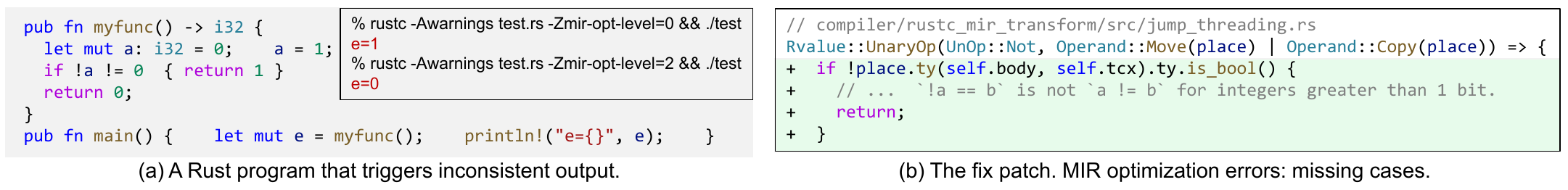}
  \caption{The example of a miscompilation bug \href{https://github.com/rust-lang/rust/issues/131195}{\textcolor{Violet}{(Issue 131195)}} and corresponding fix patch \href{https://github.com/rust-lang/rust/pull/131201}{\textcolor{Violet}{(PR 131201)}}.}
  \label{fig:exp2}
\end{figure}

\textbf{Example.}
Figure~\ref{fig:exp2} illustrates an \textit{inconsistent output} bug, caused by a \textit{MIR optimization error}. 
The test case defines a function \texttt{myfunc}, where the variable \texttt{a} is initialized as $0$ with type \texttt{i32}. In Rust, the \texttt{!} operator performs bitwise negation on integers and logical negation on booleans. 
Therefore, applying \texttt{!} to \texttt{a} should produce a 32-bit value with all bits set to 1, equivalent to $-1$ in two's complement representation, so the expected output is $e=1$. However, under MIR optimization level 2, the actual result is $e=0$, leading to an inconsistent output.
The bug originates from the \colortexttt{mybrown}{jump\_threading} optimization pass, where \rustc incorrectly applies optimizations to non-boolean operands. The optimizer fails to differentiate between integer and boolean negation, causing incorrect jump threading in specific cases. The patch in Figure~\ref{fig:exp2} (b) resolves this by introducing a boundary check, ensuring that only boolean operands are considered for jump threading. 

\subsection{Diagnostic Issues}
\rustc generates error messages for compilation failures and warnings for potential misuse, often accompanied by corresponding fix suggestions.
Therefore, we subdivide diagnostic issues into two categories.
In total, diagnostic issues account for 19.27\%. Among them, incorrect warning/error issues account for 6.64\%, and improper fixing suggestion issues account for 12.62\%.

\textbf{\textit{Incorrect Warning/Error:}}
After the compilation, \rustc may generate warning or error messages.
Nevertheless, these messages may be inaccurate or deceptive. 

\textbf{\textit{Improper Fixing Suggestion:}}
When handling invalid programs, \rustc often provides fix suggestions, yet these may be imprecise, or there could be a more optimal recommendation. 

\textbf{Bug cause analysis.}
Diagnostic issues in \rustc primarily stem from general errors (50.00\%), including shortcomings in error handling and suggestion-matching mechanisms. Similar to ICEs, while \rustc correctly identifies that the input program is non-compilable, incomplete error handling may lead to unclear diagnostics or ineffective fix suggestions.
In addition, type system (20.69\%) and ownership-related issues (27.59\%) also contribute significantly, as imprecise error branch selection within these checkers can result in misleading or unclear messages. 
Finally, MIR optimization (1.72\%) rarely causes diagnostic issues, as it involves minimal error reporting, but fix suggestions may still be affected by transformations like incorrect dead code elimination, which can remove useful code and lead to inaccurate suggestions.

\begin{figure}[htbp]
  \centering
  \includegraphics[width = \linewidth]{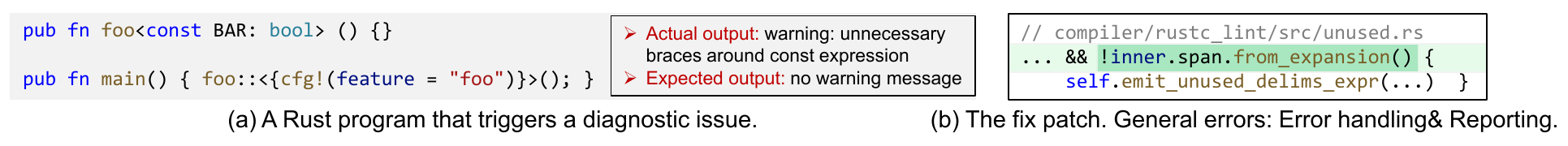}
  \caption{The example of a crash bug \href{https://github.com/rust-lang/rust/issues/104141}{\textcolor{Violet}{(Issue 104141)}} and corresponding fix patch \href{https://github.com/rust-lang/rust/pull/105515}{\textcolor{Violet}{(PR 105515)}}.}
  \label{fig:exp_warning}
\end{figure}

\textbf{Example.}
Figure~\ref{fig:exp_warning} (a) shows a code snippet that triggers a warning message suggesting the removal of unnecessary braces around a const expression. However, this warning is not correct, and applying it leads to a compilation failure. 
Because the braces are required for const generics combining with the \texttt{cfg!} macro.
This bug is classified as an error in \textit{error handling and reporting}, which falls under \textit{general errors}. Figure~\ref{fig:exp_warning} (b) presents the corresponding fix, refining the linting process to exclude edge cases involving macros in const generics.

\subsection{Misoptimization}
Misoptimization bugs in \rustc occur during the optimization phase.
While the final execution results may be correct, the MIR generated by \rustc may not match the expected optimizations. 
Besides, intermediate compilation stages can introduce inefficiencies or subtle issues, affecting soundness or performance.
In total, misoptimization issues account for 4.98\% of all. 

\textbf{Bug cause analysis.}
Misoptimization bugs in \rustc are primarily caused by MIR optimization errors (46.67\%) and general errors (40.00\%). 
Unlike crash or diagnostic issues, misoptimizations are not explicitly detected but instead manifest as deviations in the generated MIR from expected behavior. 
These issues often stem from flaws in MIR optimization algorithms or unhandled corner cases in \rustc.
Additionally, the type system (6.67\%) and ownership-related issues (6.67\%) contribute to a smaller portion of misoptimizations. In these cases, incorrect analyses can propagate errors into MIR lowering, leading to unintended transformations in the optimized code.

\begin{figure}[htbp]
  \centering
  \includegraphics[width = \linewidth]{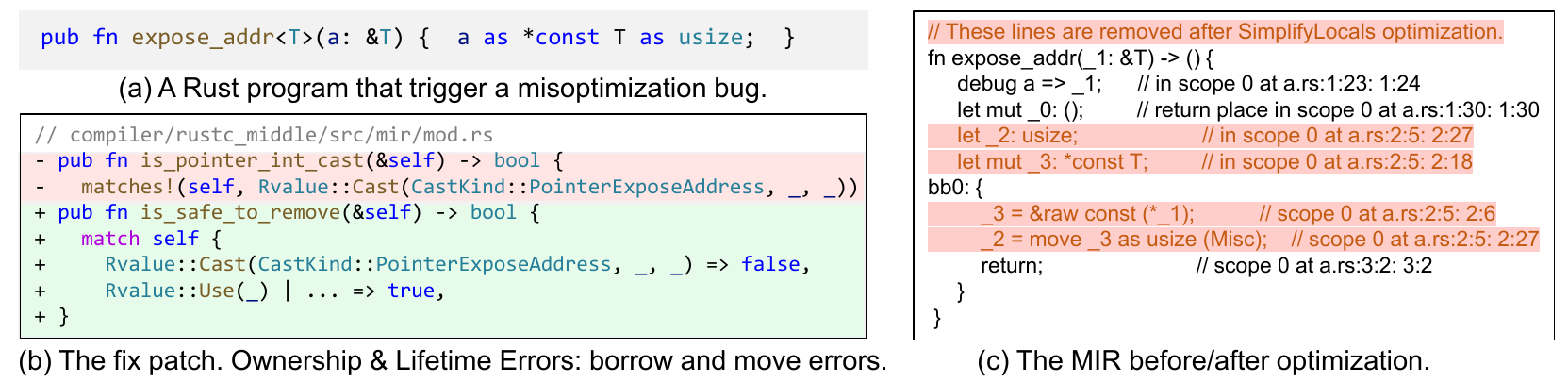}
  \caption{The example of a misoptimization bug \href{https://github.com/rust-lang/rust/issues/97421}{\textcolor{Violet}{(Issue 97421)}} and corresponding fix patch \href{https://github.com/rust-lang/rust/pull/97597}{\textcolor{Violet}{(PR 97597)}}.}
  \label{fig:exp3}
\end{figure}

\textbf{Example.}
Figure~\ref{fig:exp3} illustrates a \textit{misoptimization} caused by an \textit{ownership \& lifetime error}, specifically a \textit{borrow and move error}. 
The issue arises in the \colortexttt{mybrown}{SimplifyLocals} optimization pass, which removes unused variables and redundant code at the MIR level. 
As shown in Figure~\ref{fig:exp3} (c), the optimizer incorrectly eliminates \texttt{\_2} (a \texttt{usize} variable) and \texttt{\_3} (a raw pointer of type \texttt{const T}), highlighted in red. 
Under Rust's strict provenance model, pointer-to-integer conversions must retain provenance information, as they encode the pointer's origin. While the program may still compile and execute, a deeper MIR-level analysis reveals deviations from expected behavior.
The patch in Figure~\ref{fig:exp3} (b) fixes this by introducing stricter validation for pointer-to-integer casts, ensuring they are preserved when necessary. 

\section{RQ3: Test Case Characteristics}\label{sec:rq3_characteristics}
In this section, we analyze the characteristics of the bug-revealing test cases. 
By studying the characteristics and properties of these test cases, we can identify specific aspects of Rust that contribute to \rustc bugs, offering guidance for test case design.

A test case consists of a Rust program and a compilation command, which we collect from each issue. 
If both original and reduced test cases are reported, we collect them separately. 
If only a single test case is reported, it is categorized as an original test case. 
If a minimized version is included in the corresponding comment, it is collected as a reduced test case; otherwise, we record the reduced case as the same as the original case. 
Additionally, if a bug is reproducible only within a separate Rust project, we exclude it, as the excessive presence of unrelated elements may obscure the test case characteristics.
Among the 301 valid bugs, we have collected 276 original test cases and 293 reduced test cases.
The number of original test cases is less than the reduced ones because some original cases are separate projects, which we do not collect, yet their reduced versions are included.
There are 8 issues without test cases, due to: (1) unavailable external links, (2) separate projects, and (3) test cases with only compilation commands, not executable programs.
To analyze the characteristics of test cases, we convert the reduced test cases into an AST and count the node types that reflect syntactic structures.
Specifically, we use \textit{syn} library~\cite{syncrate62:online} to parse the AST and extract the occurrences of \textit{item} and \textit{type} nodes.
The \textit{item} nodes represent top-level constructs in Rust, such as functions, structs, traits, and enums, which define the overall structure of a program.
The \textit{type} nodes capture the different kinds of types in Rust, including primitives, references, and more complex types like trait objects, providing insights into how values are represented and manipulated in the code.
Excluding a few cases where no test case is provided or where severe syntax errors prevent generating an AST, we collect a total of $293$ test cases and $271$ corresponding ASTs.
Since \textit{syn} library cannot parse Rust code fragments without a main function, we manually supplement such snippets. 
If the test case only defines items like functions or structs, we append an empty main function. 
Otherwise, if it contains statements, we wrap the entire snippet in a main function.
Additionally, we also analyze features triggering \rustc bugs from various perspectives, including unstable features, compilation flags, built-in traits, and other keywords or APIs.
For the compilation command, we identify the most frequently used commands and their usage frequency.

\begin{wraptable}{r}{0.4\textwidth}
    \small
    \centering
    \caption{Statistics on test case sizes: lines of code (LoC).}
    \label{tab:loc}
    \setlength{\tabcolsep}{1mm}{
    \begin{tabular}{ccccc}
    \hline
    & mean  & median & min & max \\ \hline
    Original tests & 17.83 & 12     & 2   & 346 \\
    Reduced tests  & 14.17 & 11     & 2   & 123 \\
    \hline
    \end{tabular}}
\end{wraptable}

Table~\ref{tab:loc} presents some general statistics on test cases. 
The average size of original test cases is 17.83 lines of code (LoC), while the median is 12 LoC. 
Since not every test case has a reduced version, the average LoC for reduced test cases is 14.17, with a median of 11. The difference from the original test cases is not significant, although the maximum LoC has decreased from 346 to 123. Based on these statistics, we could infer that \rustc bugs are mainly triggered by small fragments of code.

\begin{table}[htbp]
\footnotesize
\renewcommand\arraystretch{0.9}
\caption{Summary of AST node types and their occurrence across test cases.}
\label{tab:item_type}
\setlength{\tabcolsep}{1.4mm}{
\begin{tabular}{ccccc||ccccc}
\toprule
\textbf{Item} & \textbf{Total} & \textbf{Prevalence} & \textbf{\begin{tabular}[c]{@{}c@{}}Mean \\ per File\end{tabular}} & \textbf{\begin{tabular}[c]{@{}c@{}}Max \\ per File\end{tabular}} & \textbf{Type} & \textbf{Total} & \textbf{Prevalence} & \textbf{\begin{tabular}[c]{@{}c@{}}Mean \\ per File\end{tabular}} & \textbf{\begin{tabular}[c]{@{}c@{}}Max \\ per File\end{tabular}} \\ \hline
\rowcolor[HTML]{EFEFEF}\textbf{Function}      & 524            & 100.00\%            & 1.93                   & 8                     & \textbf{Path}          & 1262           & 88.19\%             & 5.28                   & 41                    \\
\textbf{Struct}        & 130            & 37.64\%             & 1.27                   & 4                     & \textbf{Reference}     & 276            & 43.91\%             & 2.32                   & 10                    \\
\rowcolor[HTML]{EFEFEF}\textbf{Impl}          & 157            & 37.64\%             & 1.54                   & 6                     & \textbf{Tuple}         & 161            & 30.26\%             & 1.96                   & 8                     \\
\textbf{Trait}         & 144            & 34.32\%             & 1.55                   & 6                     & \textbf{Impl Trait}     & 87             & 20.66\%             & 1.55                   & 10                    \\
\rowcolor[HTML]{EFEFEF}\textbf{Use}           & 64             & 20.30\%             & 1.16                   & 3                     & \textbf{Array}         & 55             & 11.44\%             & 1.77                   & 10                    \\
\textbf{Type}          & 29             & 7.38\%              & 1.45                   & 6                     & \textbf{Trait Object}   & 49             & 10.70\%             & 1.69                   & 3                     \\
\rowcolor[HTML]{EFEFEF}\textbf{Enum}          & 8              & 2.95\%              & 1                      & 1                     & \textbf{Ptr}           & 35             & 7.75\%              & 1.67                   & 4                     \\
\textbf{Macro}         & 11             & 2.95\%              & 1.38                   & 2                     & \textbf{Infer}         & 18             & 4.80\%              & 1.38                   & 2                     \\
\rowcolor[HTML]{EFEFEF}\textbf{Extern Crate}   & 7              & 2.58\%              & 1                      & 1                     & \textbf{BareFn}        & 21             & 4.06\%              & 1.91                   & 5                     \\
\textbf{Static}        & 7              & 2.21\%              & 1.17                   & 2                     & \textbf{Slice}         & 13             & 2.95\%              & 1.63                   & 3                     \\
\rowcolor[HTML]{EFEFEF}\textbf{Mod}           & 8              & 1.85\%              & 1.6                    & 3                     & \textbf{Never}         & 1              & 0.37\%              & 1                      & 1                     \\
\textbf{Const}         & 5              & 1.85\%              & 1                      & 1                     & \textbf{Paren}         & 1              & 0.37\%              & 1                      & 1                     \\
\rowcolor[HTML]{EFEFEF}\textbf{Verbatim}      & 4              & 1.11\%              & 1.33                   & 2                     & \textbf{Group}         & 0              & 0.00\%              & 0                      & 0                     \\
\textbf{Foreign Mod}    & 2              & 0.74\%              & 1                      & 1                     & \textbf{Macro}         & 0              & 0.00\%              & 0                      & 0                     \\
\rowcolor[HTML]{EFEFEF}\textbf{Trait Alias}    & 1              & 0.37\%              & 1                      & 1                     & \textbf{Verbatim}      & 0              & 0.00\%              & 0                      & 0                     \\
\textbf{Union}         & 0              & 0.00\%              & 0                      & 0                     &               &                &                     &                        &          \\ \bottomrule            
\end{tabular}}
\end{table}

Table~\ref{tab:item_type} presents the distribution of \textit{Item} and \textit{Type} nodes across test cases. 
"\textit{Total}" is the total occurrences of each node, "\textit{Prevalence}" is the percentage of test cases containing the node, "\textit{Mean per File}" is its average occurrences per file, and "\textit{Max per File}" is the highest count in a single file.
Among \textit{Item} nodes, \textit{function} is the most common node, followed by \textit{struct}, \textit{impl}, and \textit{trait}, accounting for around 35\%. These nodes often appear multiple times per file, suggesting that a test case usually defines several custom data structures. 
About 20\% of test cases contain \textit{use} statements, mainly for standard library imports and some third-party dependencies.
\textit{Type}, representing custom types like type aliases, appears in nearly 10\%, common in Rust’s trait-based generics for abstract and reusable code.
Among data type nodes, \textit{Path} is the most frequent, appearing in 88.19\% of test cases. 
It represents the fully qualified name of types, e.g., \texttt{Vec<i32>}, \texttt{std::fs::File}. \textit{Reference} is the second most common type, appearing in 43.91\% of cases, reflecting Rust's ownership and borrowing system. 
This is also linked to \textit{ptr} (raw pointers), which bypass safety checks in advanced use cases. 
Trait-related types such as \textit{Impl Trait} (20.66\%) and \textit{Trait Object} (10.70\%) support compile-time and runtime polymorphism, respectively.

The other features triggering rustc bugs are listed in Table~\ref{tab:frequent_features_flags}.
Around 25\% of the test cases involved unstable features, while about 20\% required specific compilation flags.
In total, we identified 42 distinct unstable features and 41 different compilation flags. 
Many frequently used unstable features are applied to support advanced trait usages.
The \colortexttt{mybrown}{generic\_const\_exprs} feature (17.81\%) allows constant expressions in generic parameters, enabling more flexible compile-time computations. 
The \colortexttt{mybrown}{type\_alias\_impl\_trait} feature (15.07\%) simplifies complex trait bounds by allowing type aliases with \colortexttt{mybrown}{impl Trait}, making generic code more concise. 
Meanwhile, the \colortexttt{mybrown}{const\_trait\_impl} feature (4.11\%) enables trait implementations in constant contexts, further extending Rust's compile-time capabilities.
These features enhance Rust's type system but also introduce complexity to trait resolution, type inference, and constant evaluation. 
The interplay of traits, generics, and compile-time computation boosts expressiveness while increasing the edge cases \rustc must handle. 
Unstable trait-related features often reveal subtle issues in type checking, trait coherence, and monomorphization. 
Consequently, testing \rustc becomes more challenging, as ensuring soundness while supporting richer abstractions demands rigorous validation against an increasingly intricate trait system.
The other two unstable features are primarily related to low-level optimizations.
The \colortexttt{mybrown}{core\_intrinsics} feature (12.33\%) provides direct access to compiler intrinsics for performance-critical operations.
The \colortexttt{mybrown}{custom\_mir} feature (10.96\%) allows custom transformations on MIR, enabling experimental optimizations and analysis.
The most common compilation flags are related to optimization. 
The \colortexttt{mybrown}{-Zmir-opt-level=X} (45.61\%) and \colortexttt{mybrown}{-Copt-level=X} flag (14.04\%) controls MIR and LLVM optimizations, respectively. 
The \colortexttt{mybrown}{-Zmir-enable-passes=+X} flag (15.79\%) enables specific MIR passes. 
The \colortexttt{mybrown}{+nightly} flag (14.04\%) specifies the nightly \rustc version, and \colortexttt{mybrown}{-edition=X} specifies the Rust edition.

\begin{table}[htbp]
\footnotesize
\caption{The five most frequent unstable features and compilation flags required by test cases.}
\label{tab:frequent_features_flags}
\setlength{\tabcolsep}{0.5mm}{
\begin{tabular}{cc|cc|cc|cc}
\toprule
\multicolumn{2}{c|}{\textbf{Most frequent unstable features}}   & \multicolumn{2}{c|}{\textbf{Most frequent compile flags}}
& \multicolumn{2}{c|}{\textbf{Most frequent traits}}
& \multicolumn{2}{c}{\textbf{Other features}}
\\
\textbf{Feature}                & \textbf{Occ (\%)} 
& \textbf{Flag}              & \textbf{Occ (\%)}    
& \textbf{Trait}              & \textbf{Occ (\%)} 
& \textbf{Feature}              & \textbf{Occ (\%)} \\ \hline
\#!{[}feature(generic\_const\_exprs){]}    & 17.81\%           & -Zmir-opt-level=X               & 45.61\%
& (?)Sized  & 49.23\%
& lifetimes & 34.55\%
\\
\#!{[}feature(type\_alias\_impl\_trait){]} & 15.07\%           & -Zmir-enable-passes=+X          & 15.79\%
& FnOnce & 12.31\%
& std API  & 18.60\%
\\
\#!{[}feature(core\_intrinsics){]}         & 12.33\%           & -Copt-level=X                   & 14.04\%
& Iterator & 7.69\%
& \texttt{dyn}  & 9.97\%
\\
\#!{[}feature(custom\_mir){]}              & 10.96\%           & +nightly                        & 14.04\%
& Copy & 6.15\%
& \texttt{async}  & 7.31\%
\\
\#!{[}feature(const\_trait\_impl){]}       & 4.11\%            & --edition=X                     & 12.28\%
& FnMut & 4.72\%
& core API  & 6.31\%
\\ \hline
\textbf{Total: 73}        & \textbf{24.25\%}           & \textbf{Total: 57}        & \textbf{18.94\%}  &  \textbf{Total: 65} &  \textbf{21.59\%} & -& -\\ \bottomrule
\end{tabular}}
\end{table}

Given the frequent occurrence of traits in test cases, we further analyze the usage of built-in traits, which can increase test case complexity, as demonstrated by the examples in Section~\ref{subsec:type_system_errors} and Section~\ref{subsec:ownership_errors}.
Test cases involving at least one built-in trait account for 21.59\% of all cases. 
Additionally, 18.60\% of cases import standard library traits (\colortexttt{mybrown}{use std}), and 6.31\% use core library traits (\colortexttt{mybrown}{use core}). 
This suggests that the flexible use of Rust's built-in traits contributes to triggering \rustc bugs.
Table~~\ref{tab:frequent_features_flags} shows the five most frequently used build-in traits.
The \colortexttt{mybrown}{Sized} trait (49.23\%) ensures a type has a known size at compile time, while \colortexttt{mybrown}{?Sized} allows dynamically sized types like \colortexttt{mybrown}{str} and \colortexttt{mybrown}{dyn Trait}. 
The \colortexttt{mybrown}{FnOnce} trait (12.31\%) applies to types callable at most once, typically due to ownership constraints. 
The \colortexttt{mybrown}{Iterator} trait (7.69\%) enables value generation, whereas \colortexttt{mybrown}{Copy} (6.15\%) allows duplication via bitwise copying instead of moves. The \colortexttt{mybrown}{FnMut} trait (4.72\%) permits multiple calls, modifying the captured environment each time.
For other language features, lifetimes play a crucial role, with 34.55\% of test cases using lifetime annotations. Additionally, the usage of \colortexttt{mybrown}{dyn} (for dynamic dispatch via trait objects) and \colortexttt{mybrown}{async} (for asynchronous programming) also contributes to detecting \rustc bugs.

\section{RQ4: Status of Existing Techniques}\label{sec:rq4_tools}
A major concern for developers is how to automate the testing and verification of \rustc as it evolves.
Several \rustc-specific testing tools have been proposed by the Rust community and academia, and they differ in program generation and testing methods. This section reviews existing automated techniques for finding \rustc bugs.

\textbf{Analysis Method.}
Table~\ref{tab:tools} lists the selected testing tools, their first release time, program generation approaches, supported features, and testing methods.
In the Rust community, several individual projects have been developed to perform fuzz testing on \rustc. 
Fuzz-rustc~\cite{dwrensha28:online} adapts LibFuzzer~\cite{libFuzze34:online} into a custom script to systematically mutate input byte stream and uncover crashes in \rustc. 
Tree-splicer~\cite{langston76:online} constructs new test cases by recombining ASTs extracted from existing programs, though it is constrained by the structures present in its seed inputs and often produces syntactically invalid programs. 
ICEMaker~\cite{matthias37:online}, the most widely used fuzzing tool, combines elements of both Fuzz-rustc and Tree-splicer, leveraging iterative mutations and employing tools like Miri~\cite{rustlang28:online} and Clippy~\cite{rustlang22:online} to analyze generated programs. 
In academia, several tools have been developed to test \rustc by generating Rust programs using different methodologies.
RustSmith~\cite{RustSmith} constructs ASTs that conform to Rust's grammar, ensuring syntactically valid programs, and uses differential testing to detect inconsistencies across \rustc versions or optimization levels. 
Rustlantis~\cite{rustlantis} generates custom MIRs via the \texttt{mir!()} macro, making it effective at detecting bugs in MIR-based optimizations. 
Rust-twins~\cite{yang2024rusttwins} employs differential testing by generating semantically equivalent programs using macros and comparing their HIRs and MIRs, aided by Large Language Models (LLMs) for generation. 
Typecheck-fuzzer, an early work by Dewey et al.~\cite{dewey2015fuzzing}, uses Constraint Logic Programming (CLP) to generate well-typed programs and uncover type-checking bugs in Rust's type system. 

\begin{table}[htbp]
\footnotesize
\caption{Information and statistical results of existing tools for detecting \rustc bugs. (Validity: \protect\solidcirc\ indicates all of the generated programs are valid, \protect\halfsolidcirc\ indicates approximately half of the generated programs are valid, and \protect\threefourhollowcirc\ indicates the generated programs are mostly invalid. Support features: \protect\hollowcirc\ means unsupported, \protect\solidcirc\ means fully supported, and \protect\halfsolidcirc\ means partially supported for specific features.)}
\label{tab:tools}
\begin{threeparttable}
\setlength{\tabcolsep}{0.3mm}{
\begin{tabular}{ccc|ccc|ccc|ccc}
\toprule
 &    \multirow{2}{*}{\textbf{Tool}}          &     \textbf{First}          & \multicolumn{3}{c|}{\textbf{Program generation approaches}}     & \multicolumn{3}{c|}{\textbf{Supported features}}          &    \textbf{Testing}                  &  \textbf{\#Reported} & \textbf{\#Tested}     \\ \cline{4-9}
&    & \textbf{Release$^1$}  & \textbf{Method}              & \textbf{Representation} & \textbf{Validity} & \textbf{Unstable} & \textbf{Flag} &  \textbf{API} & \textbf{Method}       & \textbf{Bugs}$^2$ & \textbf{Bugs} \\ \midrule
\multirow{3}{*}{\tiny\begin{sideways}Community\end{sideways}}  & Fuzz-rustc   & 2019-07       & mutation            & Byte Stream    & \threefourhollowcirc     & \halfsolidcirc                & \hollowcirc           & \halfsolidcirc          & Fuzzing              & 49 &  1  \\
& Tree-splicer & 2023-03       & splicing            & AST            & \threefourhollowcirc     & \hollowcirc                & \hollowcirc           & \halfsolidcirc           & Fuzzing              & 27   & 0 \\
 & ICEMaker     & 2020-12       & mutation & AST            & \threefourhollowcirc     & \halfsolidcirc                & \solidcirc           & \halfsolidcirc           & Fuzzing              & 873  & 0 \\ \midrule
\multirow{4}{*}{\begin{sideways}Academia\end{sideways}}  & RustSmith    & 2022-04       & Rule-based          & AST            & \solidcirc            & \hollowcirc                & \solidcirc           & \hollowcirc           & Differential & 3  &  0  \\
& Rustlantis   & 2023-01       & Rule-based          & MIR            & \solidcirc            & \halfsolidcirc              & \halfsolidcirc         & \hollowcirc           & Differential & 8 & 0    \\
& Rust-twins   & 2024-10       & LLM-based           & Rust code          & \halfsolidcirc      & \solidcirc                & \solidcirc           & \solidcirc           & Differential & 8  & 2   \\
& CLP-Fuzzer   & 2015-10       & Rule-based          & Rust  code         & \solidcirc           & \hollowcirc                & \hollowcirc           & \hollowcirc           & Fuzzing              & 14  & - \\ \bottomrule
\end{tabular}}
\begin{tablenotes}
\footnotesize
\item[1] For tools proposed in academic papers without open-source availability, we document the publication date of the paper, the actual tool development likely preceded this date.
\item[2] The bugs detected by these tools do not fully align with our dataset. For community-sourced tools, we use their official bug statistics (up to March 3, 2025), and for paper-proposed tools, we record the data from their publications.
\end{tablenotes}
\end{threeparttable}
\end{table}

To investigate the performance of these tools, we conduct a \textit{two-step} analysis. 
In the first step, we identify which \rustc bugs in our dataset fall within the tool's capability scope by matching the user names of the issue submitters. 
If the submitter is a tool developer, we attribute the bug detection to that tool. If the submitter belongs to the Rust development team, we classify the issue as detected by a Rust team member; otherwise, it is reported by a Rust user.
In addition, we examine all these open-source tools and review their corresponding papers to understand their techniques.
In the second step, we run each tool for 12 hours to test a specific historical version of \rustc (\href{https://github.com/rust-lang/rust/releases/tag/1.58.0}{v1.58.0}), recording the number and types of detected bugs.
For tools that require seed programs, we use the official test suite of the \rustc being tested. Since the official test suite includes many cases expected to trigger ICE, we exclude these cases and apply the remaining ones as the seed set.
After excluding these cases, there are a total of $6,876$ test cases.
We did not run CLP-Fuzzer because the code link is no longer available and it was tested on an early 1.0-alpha version of \rustc, which is very different from modern \rustc.
We follow the default setup (e.g., verification commands, LLM settings) of each tool in our experiment, and all experiments are conducted in the same environment.

\textbf{Analysis Results.}
In Table~\ref{tab:tools}, the "\textit{Validity}" column indicates whether the generated program can be successfully compiled. 
Community-developed fuzzers often produce invalid programs due to the randomness of program generation and the coarse-grained nature of their mutation and splicing rules. 
In contrast, academic research tends to focus more on generating valid programs, which is particularly useful for uncovering deeper \rustc bugs, such as miscompilations and misoptimizations.
The "\textit{Supported features}" column shows which high-frequency features summarized in Section~\ref{sec:rq3_characteristics} are supported by each tool.
Community-developed fuzzers rarely provide explicit support for unstable features and the std/core API, relying instead on seed programs. 
If present in seeds, these features may be incorporated during mutation. 
However, due to the lack of semantic awareness, generated test cases may declare an unstable feature or API without actually exercising its functionality, limiting their effectiveness in systematically testing such features.
Academic research often explores various compilation flag combinations, which is particularly beneficial for differential testing, yet rarely supports unstable features and APIs explicitly. 
Among these four tools, only Rust-twins fully supports them, leveraging LLMs for code generation. 
Rustlantis supports a few unstable features for custom MIR and low-level optimizations, while RustSmith and CLP-Fuzzer overlook them.
The column "\textit{\#Tested Bugs}" in Table~\ref{tab:tools} shows the number of bugs detected after we run each tool for 12 hours. 
Among them, Fuzz-rustc found 1 bug, and Rust-twins identified 2 bugs, all of which are ICEs.
Upon our careful check, the bug found by Fuzz-rustc was previously submitted by ICEMaker\footnote{\url{https://github.com/rust-lang/rust/issues/114920}} and is still in an open state, which seems that it has not been actively maintained or verified since then.
The two ICEs discovered by Rust-twins are duplicates, with identical error messages and root causes. They overlap with another issue submitted by ICEMaker in the past\footnote{\url{https://github.com/rust-lang/rust/issues/123950}}, which was closed after Rust developers determined it to be intentional behavior.
From the detection results, ICE is the most easily triggered and detected bug symptom. We believe that the effectiveness of testing tools may be influenced by a longer testing time and the quality of seed programs.

\begin{figure}[t]
    \centering
    \subfigure[Distribution of bug symptoms across existing tools.]{
        \label{subfig:symptoms_tools}
        \includegraphics[width=0.48\linewidth]{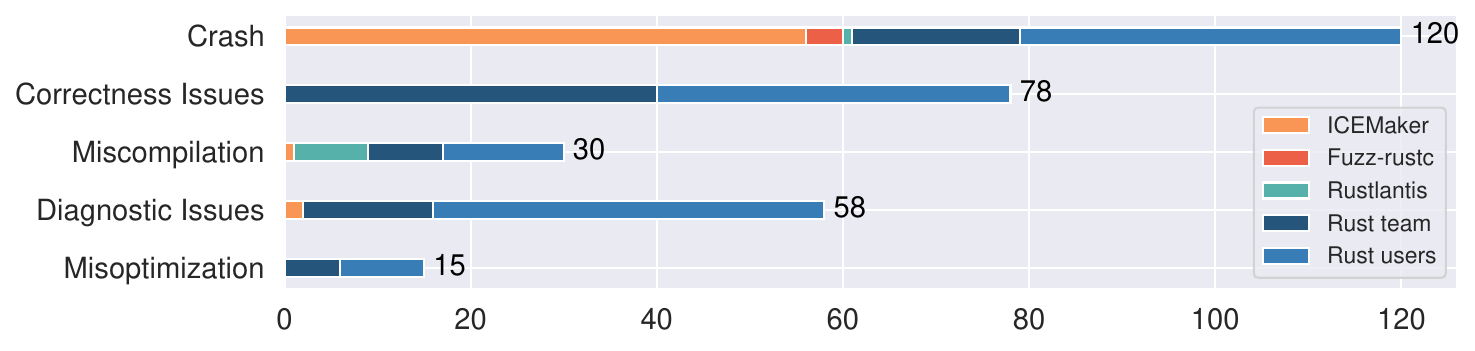}
    }
    \subfigure[Distribution of bug causes across testing tools.]{
        \label{subfig:cause_tools}
        \includegraphics[width=0.48\linewidth]{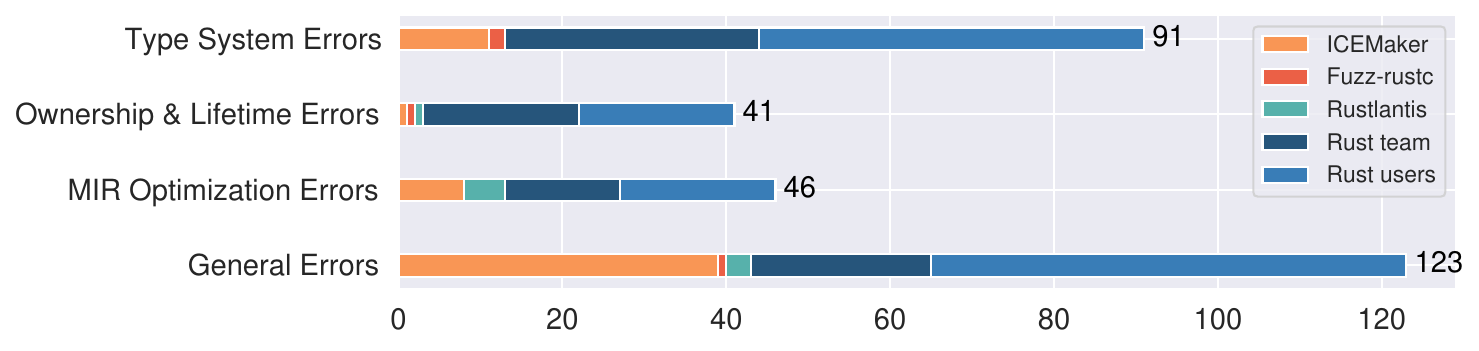}
    }
    \caption{Distribution of bug causes and symptoms across existing tools.}
    \label{fig:tools_symptom_cause}
\end{figure}

Figure~\ref{fig:tools_symptom_cause} presents the detection status of the 301 valid bugs we collected across different tools. 
Among them, ICEMaker, Fuzz-rustc, and Rustlantis have detected some bugs, while issues submitted by other tools are not included in our collected bug list.
This discrepancy may be due to some tools submitting issues beyond our dataset collection timeframe, their reported issues lacking the labels we collected, or their submitted issues remaining open and thus not included.
As shown in Figure~\ref{subfig:symptoms_tools}, ICEMaker and Fuzz-rustc, as fuzzing tools, are capable of detecting crash bugs, while Rustlantis specializes in identifying miscompilation bugs. 
Additionally, ICEMaker has also detected a few miscompilation and diagnostic issues, thanks to its more comprehensive verification approach, which leverages a wider range of compilation flags and integrates tools like Miri to detect UB. 
However, as Figure~\ref{subfig:cause_tools} indicates, the majority of the bugs detected by ICEMaker are caused by general errors, rather than Rust-specific issues, highlighting ICEMaker's ability to uncover corner cases within \rustc.
An interesting observation from Figure~\ref{subfig:symptoms_tools} is that existing tools fail to detect correctness issues and misoptimizations. This highlights the limitations of current tools in identifying deep \rustc bugs, which require a deep understanding of Rust's language rules and extensive experience.

\section{Implications and Discussion}\label{sec:discussion}
This section analyzes the conclusions and findings derived from our study, and we further discuss potential future research directions for \rustc, and threats to the validity.

\subsection{Findings}

\begin{enumerate}[label={\tmark~\textbf{Finding~\arabic*:}}, wide, labelindent=0em, itemsep=5pt, topsep=2pt]

\item \textbf{A large number of \rustc bugs in the HIR and MIR modules are caused by Rust's unique type system and lifetime model.}
As shown in Table~\ref{tab:bug_causes}, most bugs are related to the type system (30.23\%), ownership and lifetime management (13.62\%), and MIR optimizations (15.28\%). 
Unlike many compilers that lower the AST directly to LLVM IR~\cite{clang}, \rustc introduces HIR and MIR for advanced type and lifetime analysis, as well as compilation optimizations.
Figure~\ref{subfig:stages_distribution} shows that bug distributions in HIR (44.85\%) and MIR (35.22\%) are similar, underscoring frequent errors tied to Rust's advanced type and memory management mechanisms. These findings highlight the challenges posed by Rust-specific features, such as zero-cost abstractions and strict borrow checking, making these components more error-prone than general programming features. 

\item \textbf{\rustc bugs share many symptoms with other compiler bugs but also introduce unique types, such as undefined behavior in safe Rust.}
Like other compilers, \rustc experiences various compilation and runtime bugs. However, its ICE bug often causes panic with safety protection, setting it apart from other compilers where ICE typically results in segmentation faults or abnormal terminations. 
Another unique symptom is undefined behavior in safe Rust code, tied to Rust's safety guarantees. 
While performance-related bugs are absent in our analysis, this doesn't mean \rustc is free of performance issues. Rather, these issues tend to appear less frequently in Rust-specific issues or may be categorized as misoptimizations related to code efficiency.

\item \textbf{Both valid and invalid programs are critical in triggering \rustc bugs.}
As shown in Table~\ref{tab:symptoms}, certain bug symptoms such as partial ICEs, completeness issues, and inconsistencies require valid programs to reveal, while issues related to soundness, diagnostics, and some ICEs can only be triggered by invalid programs. 
This highlights the importance of both valid and invalid programs in exposing different types of \rustc bugs.
This contrasts with the typical approaches in compiler testing, where front-end bugs are often triggered by invalid programs~\cite{chaliasos2022finding, chen2016coverage, sun2018Perses}, while optimization bugs are primarily tested with valid programs~\cite{le2014compiler, Livinskii2020random, yang2011finding}. 
It's important to note that invalid programs are not entirely outside language rules; they often contain minor errors that trigger bugs in the core components, such as trait solving and optimization. 

\item \textbf{\rustc's diagnostic module still has considerable potential for enhancement, with many issues distributed across different IR-processing modules.}
As shown in Table~\ref{tab:bug_causes}, diagnostic issues account for about 20\% of all bugs. 
Figure~\ref{subfig:hir_mir_distribution} illustrates that error reporting is scattered across different components, including HIR (14.07\%) and MIR (16.04\%), with each component having its own dedicated module for error analysis and reporting.
Moreover, gaps in these modules still exist, causing some errors to be inaccurately detected or reported.

\item \textbf{Trait-related features, unstable features, compilation flags, lifetime annotations, and standard APIs are key contributors to triggering \rustc bugs.} 
Trait-related constructs, including traits, \texttt{impl traits}, and trait objects, often appear in both item and type node types (Table~\ref{tab:frequent_features_flags}), highlighting their contribution in triggering \rustc bugs. 
From Figure~\ref{subfig:hir_mir_distribution}, trait-solving issues account for 27.41\% of HIR-related bugs, reflecting the complexity of trait components. 
Traits, frequently combined with generics and trait-bound constraints, create intricate edge cases that challenge \rustc's resolution. 
Additionally, unstable features, compilation flags, and lifetime annotations also impact the likelihood of triggering \rustc bugs. 
Applying standard APIs further exposes \rustc bugs, underscoring the interplay of multiple factors in bug occurrence.

\item \textbf{Existing \rustc testing tools are less effective at detecting non-crash bugs.}
Figure~\ref{subfig:symptoms_tools} shows that about 50\% of the crash bugs are detected by existing \rustc testing tools.
However, other issues, such as miscompilation and diagnostic errors, are largely overlooked.
The effectiveness of existing tools may be limited by their test program generation methods, for example, lacking effective guidance metrics or mutation rules. 

\end{enumerate}

\subsection{Actionable Suggestions and Takeaways}
After presenting our study results and findings, we now briefly discuss further suggestions for Rust developers, \rustc developers, and programming language researchers, respectively.

\begin{enumerate}[label={\tmark~\textbf{Suggestion~\arabic*:}}, wide, labelindent=0em, itemsep=5pt, topsep=2pt]

\item \textbf{(For Rust developers) Be cautious with unstable features and custom optimization settings.}
As shown in Table~\ref{tab:frequent_features_flags}, unstable features account for over 20\% of triggering \rustc bugs, indicating that these features may introduce flaws leading to unexpected behavior. 
Additionally, custom optimization settings, such as enabling specific MIR passes or adjusting optimization levels, can cause unintended side effects or instability.
\textit{(1) Rust developers should first execute programs with the default optimization level to check results before applying higher-level optimizations, ensuring consistency and avoiding potential \rustc bugs.
(2) Rust developers should avoid using unstable features and employ a stable version of \rustc when developing system-level software, which is beneficial for ensuring software reliability.}

\item \textbf{(For Rust developers) The suggestions provided by \rustc may be inaccurate.}
As shown in Table~\ref{tab:symptoms}, nearly 20\% of \rustc bugs are linked to the feedback provided by \rustc, including error messages and suggested fixes. 
This suggests that \rustc's diagnostic tools may not always provide accurate or effective solutions. 
\textit{If \rustc's suggestion does not resolve the issue, Rust developers should consider alternative approaches. Reporting the bug to the Rust team can also be beneficial for improving the reliability of \rustc.}

\item \textbf{(For \rustc developers) Designing testing and verification techniques for \rustc components across different IRs.}
The core process of \rustc involves HIR and MIR lowering, along with type checking, borrow checking, and optimization. 
Figure~\ref{fig:stage_cause} indicates that 44.85\% and 35.22\% of the issues occur in the modules responsible for processing HIR and MIR, respectively.
However, existing fuzzers rarely employ specialized testing techniques for these components. 
Currently, Rustlantis is the only tool capable of generating valid MIR, but it lacks support for other modules, such as type checking and lifetime analysis. 
\textit{To verify the key \rustc components, \rustc developers should generate valid HIRs and MIRs under specific constraints. For example, generating HIRs to ensure well-formedness in different scenarios, such as for build-in traits and user-defined traits.}



\item \textbf{(For \rustc developers) Investing more effort in implementing and maintaining new language features and compilation settings.}
As shown in Table~\ref{tab:frequent_features_flags}, 24.25\% of the bug-revealing test cases apply unstable features, and 18.94\% employ special compilation commands. This indicates that some less frequently used or newly proposed features still have many flaws, which should receive attention from \rustc developers.
\textit{For newly proposed unstable features or syntax rules, developers should discuss thoroughly their potential use cases and \rustc's expected behaviors in RFC meetings. This helps design diverse test cases, ultimately enhancing \rustc's reliability.}

\item \textbf{(For researchers) Building better Rust program generators that fully support Rust’s unique type system.}
Research on testing, debugging, and analyzing C/C++ compilers often relies on CSmith~\cite{yang2011finding}, a random generator that produces valid C programs covering a wide range of syntax features. For Rust, the only preliminary tool, RustSmith~\cite{RustSmith}, generates complex control flow and extensive use of variables and primitive types but has limited support for Rust's higher-level abstractions. As shown in Table~\ref{tab:bug_causes}, many \rustc bugs stem from improper handling of advanced features like traits, opaque types, and references. Additionally, Table~\ref{tab:item_type} indicates that test cases combining these abstractions are more likely to trigger bugs. 
\textit{Researchers should create a Rust program generator that supports Rust's advanced features like generics, traits, and lifetime annotations, for example, by enhancing RustSmith.}

\item  \textbf{(For researchers) Generating well-designed, both valid and invalid Rust programs to test \rustc's type system.}
Our analysis shows that over half of \rustc bugs originate from the HIR and MIR modules, particularly in type and WF checking, trait resolution, borrow checking, and MIR transformation. 
Many corner cases expose weaknesses in \rustc's type handling. 
\textit{(1) Researchers should develop Rust-specific mutation rules, such as altering lifetimes, to introduce minor errors into valid programs and generate invalid ones for detecting soundness bugs.}
\textit{(2) Researchers should synthesize test programs from real-world Rust code, which provides diverse unstable features, std API usage, lifetime annotations, and complex trait patterns that benefit for testing \rustc.}

\end{enumerate}

\subsection{Threats to Validity}

One potential threat to validity is that the selected bugs may not be representative because we only collected closed issues with fixing patches and did not consider the issues still open.
We consider that unclosed issues represent the possibility that the bug has not yet been recognized by the developer or fixed, which makes it difficult to analyze the root causes and compilation stages.
We used several methods to ensure the validity.
On the one hand, we crawled all issues that were tagged with a series of labels with Rust-specific features to make sure we didn't miss the \rustc-related bugs.
In addition, we excluded non-bug issues, such as questions and discussions during the manual review process. 
This is in line with previous bug study work~\cite{Di2017A, 10.1145/2345156.2254075, sun2016toward, chaliasos2021well}, which mainly focuses on closed issues and bug reports for further analysis.
On the other hand, we have manually checked all closed issues since the release of Rust Edition 2021, covering the entire three-year period, which allows for more accurate statistical analysis in large-scale bug research.

Another potential threat is the possibility of incorrection in our labeling results. 
To mitigate this issue, we establish criteria for classifying each label, drawing references from existing compiler bug studies and the official Rust documentation. 
Additionally, each issue is independently inspected by two co-authors and then cross-checked the results between themselves and the other co-authors to achieve consensus.
This aligns with the bug analysis approach from prior empirical studies~\cite{sun2016toward, chaliasos2021well, xie2021towards,xiong2023an}, where each bug was manually reviewed and labeled by multiple researchers.

\section{Related Work}\label{sec:related_work}
In this section, we primarily focus on two perspectives of closely related research: (1) the empirical studies of compiler bugs, and (2) the studies of Rust programs.

\subsection{Understanding Compiler Bugs}
The most relevant bug study to our work is conducted by Chaliasos et al.~\cite{chaliasos2021well}, which analyzes typing-related bugs in four JVM compilers: Java, Scala, Kotlin, and Groovy. It highlights numerous overlooked type-related bugs in JVM compilers. While some findings align with ours, the design differences between \rustc and JVM compilers are significant. Notably, Rust's use of associated functions, types, and borrow checking introduces new type-related bugs.
Another closely related study by Xia et al.~\cite{xia2023understanding} provides the first analysis of historical bugs in two Rust compilers, \rustc and Rust-GCC. However, their analysis relies solely on statistical data, such as lines of code in issues, variable counts, label classifications, and affected modules in pull requests, without delving into the \rustc's implementation details. The analysis lacks depth, for example, it fails to elucidate the symptoms and root causes of the errors within \rustc.
In contrast, our work presents the first comprehensive bug analysis specifically for \rustc, the only official and mature Rust compiler. 
We manually reviewed and annotated issues and PRs related to Rust features covering a three-year period, categorizing and quantifying their bug causes and symptoms.
Our study also examines the susceptibility of different compilation stages to bugs and compares existing testing techniques for \rustc. By offering deeper insights into \rustc's design and prevalent bugs, we aim to inform researchers and guide future improvements in Rust compiler development.

Empirical studies on compiler bugs have been conducted extensively, especially for C/C++ compilers, such as the investigation proposed by Sun et.al.~\cite{sun2016toward}, which focused on understanding compiler bugs in GCC and LLVM.
Subsequently, Zhou et al.~\cite{zhou2021empirical} conducted further research and analysis on the characteristics of optimization bugs in GCC and LLVM, providing some testing and debugging guidance for testing compilers.
Another study~\cite{xie2021towards} analyzed LLVM's tool-chain bugs, summarizing typical reasons for their interaction and their corresponding fixing commits.
Additionally, an empirical study on WebAssembly compilers~\cite{Romano2022An} investigated the bugs' lifecycle, impact, and sizes of bug-inducing inputs and bug fixes.
Unlike these works, which all focus on investigating the bug characteristics of the compiler back-end, our work is the first systematic study towards \rustc as a front-end compiler.

\subsection{Empirical Studies of Rust Programs and Testing Approaches}
Most existing studies focus on the unsafe usages of Rust, such as investigating how programmers employ unsafe Rust~\cite{astrauskas2020programmers,Zhang2023On,van2023Memory,cui2024unsafe}, the potential risks associated with unsafe code~\cite{holtervennhoff2023wouldn}, and whether Rust programs are used safely~\cite{evans2020rust}.
Zhu et al.~\cite{zhu2022learning} analyzed the difficulty of understanding, application, and challenges associated with Rust safety rules.
Xu et al.~\cite{xu2021memory} conducted an in-depth analysis of Rust CVEs, exploring bugs related to memory safety.
Qin et al.~\cite{qin2020understanding} conducted research on memory and thread safety issues in real Rust programs. 
Zheng et.al.~\cite{Zheng2023A} performed an investigation into the security risks in the Rust ecosystem, discussing the characteristics of the vulnerabilities in Rust programs.
Different from existing empirical studies of Rust programs, we propose the first systematic bug study of the Rust compiler, which is resilient to memory safety issues and has unique intermediate representations designed for type checking and borrow checking.
We not only present the symptoms and root causes of \rustc bugs but also analyze the syntactic features that are prone to triggering \rustc bugs, along with the modules within \rustc that are more prone to bugs.

With Rust's powerful type system and memory management model, some research has been conducted on the testing and verification of Rust programs.
For instance, SyRust~\cite{Takashima2021SyRust} automatically generates Rust programs to effectively test Rust libraries. Verus~\cite{Lattuada2023Verus} is an SMT-based verifier for Rust programs, while Aeneas~\cite{Ho2022Aeneas} translates lightweight functions for verification. RustHornBelt~\cite{Matsushita2022RustHornBelt} employs a semantic model to check Rust's soundness. Additionally, some approaches~\cite{Wolff2021Modular, Astrauskas2019Leveraging} leverage Rust's type system for verification.
Unlike these works focusing on testing and verification, our study examines the Rust compilation process, particularly the reliability of its type-checking and borrow-checking implementations. We believe our findings can benefit compiler developers, Rust programmers, and programming language researchers while opening new directions for Rust research.

\section{Conclusion}
This paper conducts a comprehensive empirical study specifically dedicated to bugs in the Rust compiler, categorizing and analyzing bug symptoms, causes, bug-prone compilation stages, and test case characteristics. 
Many insights, suggestions, and potential research directions for testing and debugging \rustc are provided in our study. 
We found that the number of \rustc bugs caused by HIR and MIR are comparable, with errors resulting from Rust-specific analyses and checks, as well as MIR-based optimizations, accounting for the majority. 
Additionally, existing testing generation techniques for \rustc's testing tools are insufficient, and there is a lack of automated testing technologies for both correctness and misoptimization bugs in \rustc.
We expect our research to deepen the understanding of bugs in \rustc and provide guidance for \rustc's testing and development, as well as research on Rust's compilation and optimization.

\section*{Data-Availability Statement}
All source code and data for this study can be found at \url{https://anonymous.4open.science/r/rustc_bug_study-4106}.

\bibliographystyle{ACM-Reference-Format}
\bibliography{ref}


\begin{thebibliography}{61}


\ifx \showCODEN    \undefined \def \showCODEN     #1{\unskip}     \fi
\ifx \showDOI      \undefined \def \showDOI       #1{#1}\fi
\ifx \showISBNx    \undefined \def \showISBNx     #1{\unskip}     \fi
\ifx \showISBNxiii \undefined \def \showISBNxiii  #1{\unskip}     \fi
\ifx \showISSN     \undefined \def \showISSN      #1{\unskip}     \fi
\ifx \showLCCN     \undefined \def \showLCCN      #1{\unskip}     \fi
\ifx \shownote     \undefined \def \shownote      #1{#1}          \fi
\ifx \showarticletitle \undefined \def \showarticletitle #1{#1}   \fi
\ifx \showURL      \undefined \def \showURL       {\relax}        \fi
\providecommand\bibfield[2]{#2}
\providecommand\bibinfo[2]{#2}
\providecommand\natexlab[1]{#1}
\providecommand\showeprint[2][]{arXiv:#2}

\bibitem[\protect\citeauthoryear{Astrauskas, Matheja, Poli, M{\"u}ller, and Summers}{Astrauskas et~al\mbox{.}}{2020}]%
        {astrauskas2020programmers}
\bibfield{author}{\bibinfo{person}{Vytautas Astrauskas}, \bibinfo{person}{Christoph Matheja}, \bibinfo{person}{Federico Poli}, \bibinfo{person}{Peter M{\"u}ller}, {and} \bibinfo{person}{Alexander~J Summers}.} \bibinfo{year}{2020}\natexlab{}.
\newblock \showarticletitle{How do programmers use unsafe rust?}
\newblock \bibinfo{journal}{\emph{Proceedings of the ACM on Programming Languages}} \bibinfo{volume}{4}, \bibinfo{number}{OOPSLA} (\bibinfo{year}{2020}), \bibinfo{pages}{1--27}.
\newblock


\bibitem[\protect\citeauthoryear{Astrauskas, M\"{u}ller, Poli, and Summers}{Astrauskas et~al\mbox{.}}{2019}]%
        {Astrauskas2019Leveraging}
\bibfield{author}{\bibinfo{person}{Vytautas Astrauskas}, \bibinfo{person}{Peter M\"{u}ller}, \bibinfo{person}{Federico Poli}, {and} \bibinfo{person}{Alexander~J. Summers}.} \bibinfo{year}{2019}\natexlab{}.
\newblock \showarticletitle{Leveraging rust types for modular specification and verification}.
\newblock \bibinfo{journal}{\emph{Proc. ACM Program. Lang.}} \bibinfo{volume}{3}, \bibinfo{number}{OOPSLA}, Article \bibinfo{articleno}{147} (\bibinfo{date}{oct} \bibinfo{year}{2019}), \bibinfo{numpages}{30}~pages.
\newblock
\urldef\tempurl%
\url{https://doi.org/10.1145/3360573}
\showDOI{\tempurl}


\bibitem[\protect\citeauthoryear{Barrett}{Barrett}{2023}]%
        {langston76:online}
\bibfield{author}{\bibinfo{person}{Langston Barrett}.} \bibinfo{year}{2023}\natexlab{}.
\newblock \bibinfo{title}{langston-barrett/tree-splicer: Simple grammar-based test case generator}.
\newblock \bibinfo{howpublished}{\url{https://github.com/langston-barrett/tree-splicer}}.
\newblock
\newblock
\shownote{(Accessed on 12/05/2023)}.


\bibitem[\protect\citeauthoryear{Chaliasos, Sotiropoulos, Drosos, Mitropoulos, Mitropoulos, and Spinellis}{Chaliasos et~al\mbox{.}}{2021}]%
        {chaliasos2021well}
\bibfield{author}{\bibinfo{person}{Stefanos Chaliasos}, \bibinfo{person}{Thodoris Sotiropoulos}, \bibinfo{person}{Georgios-Petros Drosos}, \bibinfo{person}{Charalambos Mitropoulos}, \bibinfo{person}{Dimitris Mitropoulos}, {and} \bibinfo{person}{Diomidis Spinellis}.} \bibinfo{year}{2021}\natexlab{}.
\newblock \showarticletitle{Well-typed programs can go wrong: A study of typing-related bugs in jvm compilers}.
\newblock \bibinfo{journal}{\emph{Proceedings of the ACM on Programming Languages}} \bibinfo{volume}{5}, \bibinfo{number}{OOPSLA} (\bibinfo{year}{2021}), \bibinfo{pages}{1--30}.
\newblock


\bibitem[\protect\citeauthoryear{Chaliasos, Sotiropoulos, Spinellis, Gervais, Livshits, and Mitropoulos}{Chaliasos et~al\mbox{.}}{2022}]%
        {chaliasos2022finding}
\bibfield{author}{\bibinfo{person}{Stefanos Chaliasos}, \bibinfo{person}{Thodoris Sotiropoulos}, \bibinfo{person}{Diomidis Spinellis}, \bibinfo{person}{Arthur Gervais}, \bibinfo{person}{Benjamin Livshits}, {and} \bibinfo{person}{Dimitris Mitropoulos}.} \bibinfo{year}{2022}\natexlab{}.
\newblock \showarticletitle{Finding typing compiler bugs}. In \bibinfo{booktitle}{\emph{Proceedings of the 43rd ACM SIGPLAN International Conference on Programming Language Design and Implementation}}. \bibinfo{pages}{183--198}.
\newblock


\bibitem[\protect\citeauthoryear{Chen, Su, Sun, Su, and Zhao}{Chen et~al\mbox{.}}{2016}]%
        {chen2016coverage}
\bibfield{author}{\bibinfo{person}{Yuting Chen}, \bibinfo{person}{Ting Su}, \bibinfo{person}{Chengnian Sun}, \bibinfo{person}{Zhendong Su}, {and} \bibinfo{person}{Jianjun Zhao}.} \bibinfo{year}{2016}\natexlab{}.
\newblock \showarticletitle{Coverage-directed differential testing of JVM implementations}. In \bibinfo{booktitle}{\emph{Proceedings of the 37th ACM SIGPLAN Conference on Programming Language Design and Implementation}} (Santa Barbara, CA, USA) \emph{(\bibinfo{series}{PLDI '16})}. \bibinfo{publisher}{Association for Computing Machinery}, \bibinfo{address}{New York, NY, USA}, \bibinfo{pages}{85–99}.
\newblock
\showISBNx{9781450342612}
\urldef\tempurl%
\url{https://doi.org/10.1145/2908080.2908095}
\showDOI{\tempurl}


\bibitem[\protect\citeauthoryear{Cloudflare}{Cloudflare}{2023}]%
        {Cloudfla15:online}
\bibfield{author}{\bibinfo{person}{Cloudflare}.} \bibinfo{year}{2023}\natexlab{}.
\newblock \bibinfo{title}{Cloudflare - The Web Performance \& Security Company}.
\newblock \bibinfo{howpublished}{\url{https://www.cloudflare.com/}}.
\newblock
\newblock
\shownote{(Accessed on 12/06/2023)}.


\bibitem[\protect\citeauthoryear{Cui, Sun, Xu, and Zhou}{Cui et~al\mbox{.}}{2024}]%
        {cui2024unsafe}
\bibfield{author}{\bibinfo{person}{Mohan Cui}, \bibinfo{person}{Shuran Sun}, \bibinfo{person}{Hui Xu}, {and} \bibinfo{person}{Yangfan Zhou}.} \bibinfo{year}{2024}\natexlab{}.
\newblock \showarticletitle{Is unsafe an Achilles' Heel? A Comprehensive Study of Safety Requirements in Unsafe Rust Programming}. In \bibinfo{booktitle}{\emph{Proceedings of the IEEE/ACM 46th International Conference on Software Engineering}} (Lisbon, Portugal) \emph{(\bibinfo{series}{ICSE '24})}. \bibinfo{publisher}{Association for Computing Machinery}, \bibinfo{address}{New York, NY, USA}, Article \bibinfo{articleno}{106}, \bibinfo{numpages}{13}~pages.
\newblock
\showISBNx{9798400702174}
\urldef\tempurl%
\url{https://doi.org/10.1145/3597503.3639136}
\showDOI{\tempurl}


\bibitem[\protect\citeauthoryear{Dewey, Roesch, and Hardekopf}{Dewey et~al\mbox{.}}{2015}]%
        {dewey2015fuzzing}
\bibfield{author}{\bibinfo{person}{Kyle Dewey}, \bibinfo{person}{Jared Roesch}, {and} \bibinfo{person}{Ben Hardekopf}.} \bibinfo{year}{2015}\natexlab{}.
\newblock \showarticletitle{Fuzzing the Rust Typechecker Using CLP (T)}. In \bibinfo{booktitle}{\emph{2015 30th IEEE/ACM International Conference on Automated Software Engineering (ASE)}}. \bibinfo{pages}{482--493}.
\newblock
\urldef\tempurl%
\url{https://doi.org/10.1109/ASE.2015.65}
\showDOI{\tempurl}


\bibitem[\protect\citeauthoryear{Di~Franco, Guo, and Rubio-González}{Di~Franco et~al\mbox{.}}{2017}]%
        {Di2017A}
\bibfield{author}{\bibinfo{person}{Anthony Di~Franco}, \bibinfo{person}{Hui Guo}, {and} \bibinfo{person}{Cindy Rubio-González}.} \bibinfo{year}{2017}\natexlab{}.
\newblock \showarticletitle{A comprehensive study of real-world numerical bug characteristics}. In \bibinfo{booktitle}{\emph{2017 32nd IEEE/ACM International Conference on Automated Software Engineering (ASE)}}. \bibinfo{pages}{509--519}.
\newblock
\urldef\tempurl%
\url{https://doi.org/10.1109/ASE.2017.8115662}
\showDOI{\tempurl}


\bibitem[\protect\citeauthoryear{Drosos, Sotiropoulos, Alexopoulos, Mitropoulos, and Su}{Drosos et~al\mbox{.}}{2024}]%
        {Drosos2024When}
\bibfield{author}{\bibinfo{person}{Georgios-Petros Drosos}, \bibinfo{person}{Thodoris Sotiropoulos}, \bibinfo{person}{Georgios Alexopoulos}, \bibinfo{person}{Dimitris Mitropoulos}, {and} \bibinfo{person}{Zhendong Su}.} \bibinfo{year}{2024}\natexlab{}.
\newblock \showarticletitle{When Your Infrastructure Is a Buggy Program: Understanding Faults in Infrastructure as Code Ecosystems}.
\newblock \bibinfo{journal}{\emph{Proc. ACM Program. Lang.}} \bibinfo{volume}{8}, \bibinfo{number}{OOPSLA2}, Article \bibinfo{articleno}{359} (\bibinfo{date}{Oct.} \bibinfo{year}{2024}), \bibinfo{numpages}{31}~pages.
\newblock
\urldef\tempurl%
\url{https://doi.org/10.1145/3689799}
\showDOI{\tempurl}


\bibitem[\protect\citeauthoryear{Evans, Campbell, and Soffa}{Evans et~al\mbox{.}}{2020}]%
        {evans2020rust}
\bibfield{author}{\bibinfo{person}{Ana~Nora Evans}, \bibinfo{person}{Bradford Campbell}, {and} \bibinfo{person}{Mary~Lou Soffa}.} \bibinfo{year}{2020}\natexlab{}.
\newblock \showarticletitle{Is Rust used safely by software developers?}. In \bibinfo{booktitle}{\emph{Proceedings of the ACM/IEEE 42nd International Conference on Software Engineering}}. \bibinfo{pages}{246--257}.
\newblock


\bibitem[\protect\citeauthoryear{Ho and Protzenko}{Ho and Protzenko}{2022}]%
        {Ho2022Aeneas}
\bibfield{author}{\bibinfo{person}{Son Ho} {and} \bibinfo{person}{Jonathan Protzenko}.} \bibinfo{year}{2022}\natexlab{}.
\newblock \showarticletitle{Aeneas: Rust verification by functional translation}.
\newblock \bibinfo{journal}{\emph{Proc. ACM Program. Lang.}} \bibinfo{volume}{6}, \bibinfo{number}{ICFP}, Article \bibinfo{articleno}{116} (\bibinfo{date}{aug} \bibinfo{year}{2022}), \bibinfo{numpages}{31}~pages.
\newblock
\urldef\tempurl%
\url{https://doi.org/10.1145/3547647}
\showDOI{\tempurl}


\bibitem[\protect\citeauthoryear{H{\"o}ltervennhoff, Klostermeyer, W{\"o}hler, Acar, and Fahl}{H{\"o}ltervennhoff et~al\mbox{.}}{2023}]%
        {holtervennhoff2023wouldn}
\bibfield{author}{\bibinfo{person}{Sandra H{\"o}ltervennhoff}, \bibinfo{person}{Philip Klostermeyer}, \bibinfo{person}{Noah W{\"o}hler}, \bibinfo{person}{Yasemin Acar}, {and} \bibinfo{person}{Sascha Fahl}.} \bibinfo{year}{2023}\natexlab{}.
\newblock \showarticletitle{$\{$“I$\}$ wouldn't want my unsafe code to run my $\{$pacemaker”$\}$: An Interview Study on the Use, Comprehension, and Perceived Risks of Unsafe Rust}. In \bibinfo{booktitle}{\emph{32nd USENIX Security Symposium (USENIX Security 23)}}. \bibinfo{pages}{2509--2525}.
\newblock


\bibitem[\protect\citeauthoryear{InfoWorld}{InfoWorld}{2023}]%
        {WhiteHou52:online}
\bibfield{author}{\bibinfo{person}{InfoWorld}.} \bibinfo{year}{2023}\natexlab{}.
\newblock \bibinfo{title}{White House urges developers to dump C and C++ | InfoWorld}.
\newblock \bibinfo{howpublished}{\url{https://www.infoworld.com/article/3713203/white-house-urges-developers-to-dump-c-and-c.html}}.
\newblock
\newblock
\shownote{(Accessed on 03/17/2024)}.


\bibitem[\protect\citeauthoryear{Jin, Song, Shi, Scherpelz, and Lu}{Jin et~al\mbox{.}}{2012}]%
        {10.1145/2345156.2254075}
\bibfield{author}{\bibinfo{person}{Guoliang Jin}, \bibinfo{person}{Linhai Song}, \bibinfo{person}{Xiaoming Shi}, \bibinfo{person}{Joel Scherpelz}, {and} \bibinfo{person}{Shan Lu}.} \bibinfo{year}{2012}\natexlab{}.
\newblock \showarticletitle{Understanding and detecting real-world performance bugs}.
\newblock \bibinfo{journal}{\emph{SIGPLAN Not.}} \bibinfo{volume}{47}, \bibinfo{number}{6} (\bibinfo{date}{jun} \bibinfo{year}{2012}), \bibinfo{pages}{77–88}.
\newblock
\showISSN{0362-1340}
\urldef\tempurl%
\url{https://doi.org/10.1145/2345156.2254075}
\showDOI{\tempurl}


\bibitem[\protect\citeauthoryear{Jung, Jourdan, Krebbers, and Dreyer}{Jung et~al\mbox{.}}{2021}]%
        {jung2021safe}
\bibfield{author}{\bibinfo{person}{Ralf Jung}, \bibinfo{person}{Jacques-Henri Jourdan}, \bibinfo{person}{Robbert Krebbers}, {and} \bibinfo{person}{Derek Dreyer}.} \bibinfo{year}{2021}\natexlab{}.
\newblock \showarticletitle{Safe systems programming in Rust}.
\newblock \bibinfo{journal}{\emph{Commun. ACM}} \bibinfo{volume}{64}, \bibinfo{number}{4} (\bibinfo{year}{2021}), \bibinfo{pages}{144--152}.
\newblock


\bibitem[\protect\citeauthoryear{Klabnik and Nichols}{Klabnik and Nichols}{2023}]%
        {klabnik2023rust}
\bibfield{author}{\bibinfo{person}{Steve Klabnik} {and} \bibinfo{person}{Carol Nichols}.} \bibinfo{year}{2023}\natexlab{}.
\newblock \bibinfo{booktitle}{\emph{The Rust programming language}}.
\newblock \bibinfo{publisher}{No Starch Press}.
\newblock


\bibitem[\protect\citeauthoryear{Kruse and Finkel}{Kruse and Finkel}{2018}]%
        {clang}
\bibfield{author}{\bibinfo{person}{Michael Kruse} {and} \bibinfo{person}{Hal Finkel}.} \bibinfo{year}{2018}\natexlab{}.
\newblock \showarticletitle{User-Directed Loop-Transformations in Clang}. In \bibinfo{booktitle}{\emph{2018 IEEE/ACM 5th Workshop on the LLVM Compiler Infrastructure in HPC (LLVM-HPC)}}. \bibinfo{pages}{49--58}.
\newblock
\urldef\tempurl%
\url{https://doi.org/10.1109/LLVM-HPC.2018.8639402}
\showDOI{\tempurl}


\bibitem[\protect\citeauthoryear{Krüger}{Krüger}{2020}]%
        {matthias37:online}
\bibfield{author}{\bibinfo{person}{Matthias Krüger}.} \bibinfo{year}{2020}\natexlab{}.
\newblock \bibinfo{title}{matthiaskrgr/icemaker: automatially find crashes in the rust compiler \& tooling}.
\newblock \bibinfo{howpublished}{\url{https://github.com/matthiaskrgr/icemaker}}.
\newblock
\newblock
\shownote{(Accessed on 12/05/2023)}.


\bibitem[\protect\citeauthoryear{Lattner and Adve}{Lattner and Adve}{2004}]%
        {lattner2004llvm}
\bibfield{author}{\bibinfo{person}{Chris Lattner} {and} \bibinfo{person}{Vikram Adve}.} \bibinfo{year}{2004}\natexlab{}.
\newblock \showarticletitle{LLVM: A compilation framework for lifelong program analysis \& transformation}. In \bibinfo{booktitle}{\emph{International symposium on code generation and optimization, 2004. CGO 2004.}} IEEE, \bibinfo{pages}{75--86}.
\newblock


\bibitem[\protect\citeauthoryear{Lattuada, Hance, Cho, Brun, Subasinghe, Zhou, Howell, Parno, and Hawblitzel}{Lattuada et~al\mbox{.}}{2023}]%
        {Lattuada2023Verus}
\bibfield{author}{\bibinfo{person}{Andrea Lattuada}, \bibinfo{person}{Travis Hance}, \bibinfo{person}{Chanhee Cho}, \bibinfo{person}{Matthias Brun}, \bibinfo{person}{Isitha Subasinghe}, \bibinfo{person}{Yi Zhou}, \bibinfo{person}{Jon Howell}, \bibinfo{person}{Bryan Parno}, {and} \bibinfo{person}{Chris Hawblitzel}.} \bibinfo{year}{2023}\natexlab{}.
\newblock \showarticletitle{Verus: Verifying Rust Programs using Linear Ghost Types}.
\newblock \bibinfo{journal}{\emph{Proc. ACM Program. Lang.}} \bibinfo{volume}{7}, \bibinfo{number}{OOPSLA1}, Article \bibinfo{articleno}{85} (\bibinfo{date}{apr} \bibinfo{year}{2023}), \bibinfo{numpages}{30}~pages.
\newblock
\urldef\tempurl%
\url{https://doi.org/10.1145/3586037}
\showDOI{\tempurl}


\bibitem[\protect\citeauthoryear{Le, Afshari, and Su}{Le et~al\mbox{.}}{2014}]%
        {le2014compiler}
\bibfield{author}{\bibinfo{person}{Vu Le}, \bibinfo{person}{Mehrdad Afshari}, {and} \bibinfo{person}{Zhendong Su}.} \bibinfo{year}{2014}\natexlab{}.
\newblock \showarticletitle{Compiler validation via equivalence modulo inputs}.
\newblock \bibinfo{journal}{\emph{ACM Sigplan Notices}} \bibinfo{volume}{49}, \bibinfo{number}{6} (\bibinfo{year}{2014}), \bibinfo{pages}{216--226}.
\newblock


\bibitem[\protect\citeauthoryear{Livinskii, Babokin, and Regehr}{Livinskii et~al\mbox{.}}{2020}]%
        {Livinskii2020random}
\bibfield{author}{\bibinfo{person}{Vsevolod Livinskii}, \bibinfo{person}{Dmitry Babokin}, {and} \bibinfo{person}{John Regehr}.} \bibinfo{year}{2020}\natexlab{}.
\newblock \showarticletitle{Random testing for C and C++ compilers with YARPGen}.
\newblock \bibinfo{journal}{\emph{Proc. ACM Program. Lang.}} \bibinfo{volume}{4}, \bibinfo{number}{OOPSLA}, Article \bibinfo{articleno}{196} (\bibinfo{date}{Nov.} \bibinfo{year}{2020}), \bibinfo{numpages}{25}~pages.
\newblock
\urldef\tempurl%
\url{https://doi.org/10.1145/3428264}
\showDOI{\tempurl}


\bibitem[\protect\citeauthoryear{LLVM}{LLVM}{2023}]%
        {libFuzze34:online}
\bibfield{author}{\bibinfo{person}{LLVM}.} \bibinfo{year}{2023}\natexlab{}.
\newblock \bibinfo{title}{libFuzzer – a library for coverage-guided fuzz testing. — LLVM 18.0.0git documentation}.
\newblock \bibinfo{howpublished}{\url{https://llvm.org/docs/LibFuzzer.html}}.
\newblock
\newblock
\shownote{(Accessed on 12/09/2023)}.


\bibitem[\protect\citeauthoryear{Matsushita, Denis, Jourdan, and Dreyer}{Matsushita et~al\mbox{.}}{2022}]%
        {Matsushita2022RustHornBelt}
\bibfield{author}{\bibinfo{person}{Yusuke Matsushita}, \bibinfo{person}{Xavier Denis}, \bibinfo{person}{Jacques-Henri Jourdan}, {and} \bibinfo{person}{Derek Dreyer}.} \bibinfo{year}{2022}\natexlab{}.
\newblock \showarticletitle{RustHornBelt: a semantic foundation for functional verification of Rust programs with unsafe code}. In \bibinfo{booktitle}{\emph{Proceedings of the 43rd ACM SIGPLAN International Conference on Programming Language Design and Implementation}} (San Diego, CA, USA) \emph{(\bibinfo{series}{PLDI 2022})}. \bibinfo{publisher}{Association for Computing Machinery}, \bibinfo{address}{New York, NY, USA}, \bibinfo{pages}{841–856}.
\newblock
\showISBNx{9781450392655}
\urldef\tempurl%
\url{https://doi.org/10.1145/3519939.3523704}
\showDOI{\tempurl}


\bibitem[\protect\citeauthoryear{Miri}{Miri}{2023}]%
        {rustlang28:online}
\bibfield{author}{\bibinfo{person}{Miri}.} \bibinfo{year}{2023}\natexlab{}.
\newblock \bibinfo{title}{rust-lang/miri: An interpreter for Rust's mid-level intermediate representation}.
\newblock \bibinfo{howpublished}{\url{https://github.com/rust-lang/miri}}.
\newblock
\newblock
\shownote{(Accessed on 12/14/2023)}.


\bibitem[\protect\citeauthoryear{Qin, Chen, Yu, Song, and Zhang}{Qin et~al\mbox{.}}{2020}]%
        {qin2020understanding}
\bibfield{author}{\bibinfo{person}{Boqin Qin}, \bibinfo{person}{Yilun Chen}, \bibinfo{person}{Zeming Yu}, \bibinfo{person}{Linhai Song}, {and} \bibinfo{person}{Yiying Zhang}.} \bibinfo{year}{2020}\natexlab{}.
\newblock \showarticletitle{Understanding memory and thread safety practices and issues in real-world Rust programs}. In \bibinfo{booktitle}{\emph{Proceedings of the 41st ACM SIGPLAN Conference on Programming Language Design and Implementation}}. \bibinfo{pages}{763--779}.
\newblock


\bibitem[\protect\citeauthoryear{RedoxOS}{RedoxOS}{2023}]%
        {RedoxYou24:online}
\bibfield{author}{\bibinfo{person}{RedoxOS}.} \bibinfo{year}{2023}\natexlab{}.
\newblock \bibinfo{title}{Redox - Your Next(Gen) OS - Redox - Your Next(Gen) OS}.
\newblock \bibinfo{howpublished}{\url{https://www.redox-os.org/}}.
\newblock
\newblock
\shownote{(Accessed on 12/06/2023)}.


\bibitem[\protect\citeauthoryear{Renshaw}{Renshaw}{2019}]%
        {dwrensha28:online}
\bibfield{author}{\bibinfo{person}{David Renshaw}.} \bibinfo{year}{2019}\natexlab{}.
\newblock \bibinfo{title}{dwrensha/fuzz-rustc: setup for fuzzing the Rust compiler}.
\newblock \bibinfo{howpublished}{\url{https://github.com/dwrensha/fuzz-rustc}}.
\newblock
\newblock
\shownote{(Accessed on 12/05/2023)}.


\bibitem[\protect\citeauthoryear{Romano, Liu, Kwon, and Wang}{Romano et~al\mbox{.}}{2022}]%
        {Romano2022An}
\bibfield{author}{\bibinfo{person}{Alan Romano}, \bibinfo{person}{Xinyue Liu}, \bibinfo{person}{Yonghwi Kwon}, {and} \bibinfo{person}{Weihang Wang}.} \bibinfo{year}{2022}\natexlab{}.
\newblock \showarticletitle{An empirical study of bugs in webassembly compilers}. In \bibinfo{booktitle}{\emph{Proceedings of the 36th IEEE/ACM International Conference on Automated Software Engineering}} (Melbourne, Australia) \emph{(\bibinfo{series}{ASE '21})}. \bibinfo{publisher}{IEEE Press}, \bibinfo{pages}{42–54}.
\newblock
\showISBNx{9781665403375}
\urldef\tempurl%
\url{https://doi.org/10.1109/ASE51524.2021.9678776}
\showDOI{\tempurl}


\bibitem[\protect\citeauthoryear{Rust}{Rust}{2023a}]%
        {rustlang63:online}
\bibfield{author}{\bibinfo{person}{Rust}.} \bibinfo{year}{2023}\natexlab{a}.
\newblock \bibinfo{title}{rust-lang/rust: Empowering everyone to build reliable and efficient software.}
\newblock \bibinfo{howpublished}{\url{https://github.com/rust-lang/rust}}.
\newblock
\newblock
\shownote{(Accessed on 11/29/2023)}.


\bibitem[\protect\citeauthoryear{Rust}{Rust}{2023b}]%
        {Whatisru31:online}
\bibfield{author}{\bibinfo{person}{Rust}.} \bibinfo{year}{2023}\natexlab{b}.
\newblock \bibinfo{title}{What is rustc? - The rustc book}.
\newblock \bibinfo{howpublished}{\url{https://doc.rust-lang.org/rustc/what-is-rustc.html}}.
\newblock
\newblock
\shownote{(Accessed on 12/06/2023)}.


\bibitem[\protect\citeauthoryear{Rust}{Rust}{2025a}]%
        {Nextgent9:online}
\bibfield{author}{\bibinfo{person}{Rust}.} \bibinfo{year}{2025}\natexlab{a}.
\newblock \bibinfo{title}{Next-gen trait solving - Rust Compiler Development Guide}.
\newblock
\newblock
\urldef\tempurl%
\url{https://rustc-dev-guide.rust-lang.org/solve/trait-solving.html}
\showURL{%
\tempurl}
\newblock
\shownote{[Online; accessed 2025-03-03]}.


\bibitem[\protect\citeauthoryear{Rust}{Rust}{2025b}]%
        {Wellform75:online}
\bibfield{author}{\bibinfo{person}{Rust}.} \bibinfo{year}{2025}\natexlab{b}.
\newblock \bibinfo{title}{Well-formedness checking}.
\newblock
\newblock
\urldef\tempurl%
\url{https://rust-lang.github.io/chalk/book/clauses/wf.html}
\showURL{%
\tempurl}
\newblock
\shownote{[Online; accessed 2025-03-03]}.


\bibitem[\protect\citeauthoryear{Rust-clippy}{Rust-clippy}{2023}]%
        {rustlang22:online}
\bibfield{author}{\bibinfo{person}{Rust-clippy}.} \bibinfo{year}{2023}\natexlab{}.
\newblock \bibinfo{title}{rust-lang/rust-clippy: A bunch of lints to catch common mistakes and improve your Rust code. Book: https://doc.rust-lang.org/clippy/}.
\newblock \bibinfo{howpublished}{\url{https://github.com/rust-lang/rust-clippy}}.
\newblock
\newblock
\shownote{(Accessed on 12/14/2023)}.


\bibitem[\protect\citeauthoryear{Rust-GCC}{Rust-GCC}{2024}]%
        {RustGCCg76:online}
\bibfield{author}{\bibinfo{person}{Rust-GCC}.} \bibinfo{year}{2024}\natexlab{}.
\newblock \bibinfo{title}{Rust-GCC/gccrs: GCC Front-End for Rust}.
\newblock \bibinfo{howpublished}{\url{https://github.com/Rust-GCC/gccrs}}.
\newblock
\newblock
\shownote{(Accessed on 09/07/2024)}.


\bibitem[\protect\citeauthoryear{rust team}{rust team}{2025}]%
        {Labels·r6:online}
\bibfield{author}{\bibinfo{person}{rust team}.} \bibinfo{year}{2025}\natexlab{}.
\newblock \bibinfo{title}{Labels rust-lang/rust}.
\newblock
\newblock
\urldef\tempurl%
\url{https://github.com/rust-lang/rust/labels}
\showURL{%
\tempurl}
\newblock
\shownote{[Online; accessed 2025-01-19]}.


\bibitem[\protect\citeauthoryear{rustc-dev guide}{rustc-dev guide}{2025}]%
        {OpaqueTy48:online}
\bibfield{author}{\bibinfo{person}{rustc-dev guide}.} \bibinfo{year}{2025}\natexlab{}.
\newblock \bibinfo{title}{Opaque Types - Rust Compiler Development Guide}.
\newblock
\newblock
\urldef\tempurl%
\url{https://rustc-dev-guide.rust-lang.org/opaque-types-type-alias-impl-trait.html}
\showURL{%
\tempurl}
\newblock
\shownote{[Online; accessed 2025-03-06]}.


\bibitem[\protect\citeauthoryear{Servo}{Servo}{2023}]%
        {Servothe7:online}
\bibfield{author}{\bibinfo{person}{Servo}.} \bibinfo{year}{2023}\natexlab{}.
\newblock \bibinfo{title}{Servo, the embeddable, independent, memory-safe, modular, parallel web rendering engine}.
\newblock \bibinfo{howpublished}{\url{https://servo.org/}}.
\newblock
\newblock
\shownote{(Accessed on 12/06/2023)}.


\bibitem[\protect\citeauthoryear{Sharma, Yu, and Donaldson}{Sharma et~al\mbox{.}}{2023}]%
        {RustSmith}
\bibfield{author}{\bibinfo{person}{Mayank Sharma}, \bibinfo{person}{Pingshi Yu}, {and} \bibinfo{person}{Alastair~F. Donaldson}.} \bibinfo{year}{2023}\natexlab{}.
\newblock \showarticletitle{RustSmith: Random Differential Compiler Testing for Rust}. In \bibinfo{booktitle}{\emph{Proceedings of the 32nd ACM SIGSOFT International Symposium on Software Testing and Analysis}} \emph{(\bibinfo{series}{ISSTA 2023})}. \bibinfo{publisher}{Association for Computing Machinery}, \bibinfo{address}{New York, NY, USA}, \bibinfo{pages}{1483–1486}.
\newblock
\showISBNx{9798400702211}
\urldef\tempurl%
\url{https://doi.org/10.1145/3597926.3604919}
\showDOI{\tempurl}


\bibitem[\protect\citeauthoryear{STRATIS}{STRATIS}{2023}]%
        {StratisS70:online}
\bibfield{author}{\bibinfo{person}{STRATIS}.} \bibinfo{year}{2023}\natexlab{}.
\newblock \bibinfo{title}{Stratis Storage}.
\newblock \bibinfo{howpublished}{\url{https://stratis-storage.github.io/}}.
\newblock
\newblock
\shownote{(Accessed on 12/06/2023)}.


\bibitem[\protect\citeauthoryear{Sun, Le, Zhang, and Su}{Sun et~al\mbox{.}}{2016}]%
        {sun2016toward}
\bibfield{author}{\bibinfo{person}{Chengnian Sun}, \bibinfo{person}{Vu Le}, \bibinfo{person}{Qirun Zhang}, {and} \bibinfo{person}{Zhendong Su}.} \bibinfo{year}{2016}\natexlab{}.
\newblock \showarticletitle{Toward understanding compiler bugs in GCC and LLVM}. In \bibinfo{booktitle}{\emph{Proceedings of the 25th international symposium on software testing and analysis}}. \bibinfo{pages}{294--305}.
\newblock


\bibitem[\protect\citeauthoryear{Sun, Li, Zhang, Gu, and Su}{Sun et~al\mbox{.}}{2018}]%
        {sun2018Perses}
\bibfield{author}{\bibinfo{person}{Chengnian Sun}, \bibinfo{person}{Yuanbo Li}, \bibinfo{person}{Qirun Zhang}, \bibinfo{person}{Tianxiao Gu}, {and} \bibinfo{person}{Zhendong Su}.} \bibinfo{year}{2018}\natexlab{}.
\newblock \showarticletitle{Perses: syntax-guided program reduction}. In \bibinfo{booktitle}{\emph{Proceedings of the 40th International Conference on Software Engineering}} (Gothenburg, Sweden) \emph{(\bibinfo{series}{ICSE '18})}. \bibinfo{publisher}{Association for Computing Machinery}, \bibinfo{address}{New York, NY, USA}, \bibinfo{pages}{361–371}.
\newblock
\showISBNx{9781450356381}
\urldef\tempurl%
\url{https://doi.org/10.1145/3180155.3180236}
\showDOI{\tempurl}


\bibitem[\protect\citeauthoryear{Takashima, Martins, Jia, and P\u{a}s\u{a}reanu}{Takashima et~al\mbox{.}}{2021}]%
        {Takashima2021SyRust}
\bibfield{author}{\bibinfo{person}{Yoshiki Takashima}, \bibinfo{person}{Ruben Martins}, \bibinfo{person}{Limin Jia}, {and} \bibinfo{person}{Corina~S. P\u{a}s\u{a}reanu}.} \bibinfo{year}{2021}\natexlab{}.
\newblock \showarticletitle{SyRust: automatic testing of Rust libraries with semantic-aware program synthesis}. In \bibinfo{booktitle}{\emph{Proceedings of the 42nd ACM SIGPLAN International Conference on Programming Language Design and Implementation}} (Virtual, Canada) \emph{(\bibinfo{series}{PLDI 2021})}. \bibinfo{publisher}{Association for Computing Machinery}, \bibinfo{address}{New York, NY, USA}, \bibinfo{pages}{899–913}.
\newblock
\showISBNx{9781450383912}
\urldef\tempurl%
\url{https://doi.org/10.1145/3453483.3454084}
\showDOI{\tempurl}


\bibitem[\protect\citeauthoryear{TiKV}{TiKV}{2023}]%
        {TiKVTiKV64:online}
\bibfield{author}{\bibinfo{person}{TiKV}.} \bibinfo{year}{2023}\natexlab{}.
\newblock \bibinfo{title}{TiKV is a highly scalable, low latency, and easy to use key-value database.}
\newblock \bibinfo{howpublished}{\url{https://tikv.org/}}.
\newblock
\newblock
\shownote{(Accessed on 12/06/2023)}.


\bibitem[\protect\citeauthoryear{Tolnay}{Tolnay}{2025}]%
        {syncrate62:online}
\bibfield{author}{\bibinfo{person}{David Tolnay}.} \bibinfo{year}{2025}\natexlab{}.
\newblock \bibinfo{title}{syn - crates.io: Rust Package Registry}.
\newblock
\newblock
\urldef\tempurl%
\url{https://crates.io/crates/syn}
\showURL{%
\tempurl}
\newblock
\shownote{[Online; accessed 2025-03-03]}.


\bibitem[\protect\citeauthoryear{van Oorschot}{van Oorschot}{2023}]%
        {van2023Memory}
\bibfield{author}{\bibinfo{person}{Paul~C. van Oorschot}.} \bibinfo{year}{2023}\natexlab{}.
\newblock \showarticletitle{Memory Errors and Memory Safety: A Look at Java and Rust}.
\newblock \bibinfo{journal}{\emph{IEEE Security \& Privacy}} \bibinfo{volume}{21}, \bibinfo{number}{3} (\bibinfo{year}{2023}), \bibinfo{pages}{62--68}.
\newblock
\urldef\tempurl%
\url{https://doi.org/10.1109/MSEC.2023.3249719}
\showDOI{\tempurl}


\bibitem[\protect\citeauthoryear{Wang and Jung}{Wang and Jung}{2024}]%
        {rustlantis}
\bibfield{author}{\bibinfo{person}{Qian Wang} {and} \bibinfo{person}{Ralf Jung}.} \bibinfo{year}{2024}\natexlab{}.
\newblock \showarticletitle{Rustlantis: Randomized Differential Testing of the Rust Compiler}.
\newblock \bibinfo{journal}{\emph{Proc. ACM Program. Lang.}} \bibinfo{volume}{8}, \bibinfo{number}{OOPSLA2}, Article \bibinfo{articleno}{340} (\bibinfo{date}{Oct.} \bibinfo{year}{2024}), \bibinfo{numpages}{27}~pages.
\newblock
\urldef\tempurl%
\url{https://doi.org/10.1145/3689780}
\showDOI{\tempurl}


\bibitem[\protect\citeauthoryear{Wikimedia}{Wikimedia}{2004}]%
        {Undefine65:online}
\bibfield{author}{\bibinfo{person}{Wikimedia}.} \bibinfo{year}{2004}\natexlab{}.
\newblock \bibinfo{title}{Undefined behavior - Wikipedia}.
\newblock
\newblock
\urldef\tempurl%
\url{https://en.wikipedia.org/wiki/Undefined_behavior}
\showURL{%
\tempurl}
\newblock
\shownote{[Online; accessed 2025-03-23]}.


\bibitem[\protect\citeauthoryear{Wolff, B\'{\i}l\'{y}, Matheja, M\"{u}ller, and Summers}{Wolff et~al\mbox{.}}{2021}]%
        {Wolff2021Modular}
\bibfield{author}{\bibinfo{person}{Fabian Wolff}, \bibinfo{person}{Aurel B\'{\i}l\'{y}}, \bibinfo{person}{Christoph Matheja}, \bibinfo{person}{Peter M\"{u}ller}, {and} \bibinfo{person}{Alexander~J. Summers}.} \bibinfo{year}{2021}\natexlab{}.
\newblock \showarticletitle{Modular specification and verification of closures in Rust}.
\newblock \bibinfo{journal}{\emph{Proc. ACM Program. Lang.}} \bibinfo{volume}{5}, \bibinfo{number}{OOPSLA}, Article \bibinfo{articleno}{145} (\bibinfo{date}{oct} \bibinfo{year}{2021}), \bibinfo{numpages}{29}~pages.
\newblock
\urldef\tempurl%
\url{https://doi.org/10.1145/3485522}
\showDOI{\tempurl}


\bibitem[\protect\citeauthoryear{Xia, Feng, and Shi}{Xia et~al\mbox{.}}{2023}]%
        {xia2023understanding}
\bibfield{author}{\bibinfo{person}{Xinmeng Xia}, \bibinfo{person}{Yang Feng}, {and} \bibinfo{person}{Qingkai Shi}.} \bibinfo{year}{2023}\natexlab{}.
\newblock \showarticletitle{Understanding Bugs in Rust Compilers}. In \bibinfo{booktitle}{\emph{2023 IEEE 23rd International Conference on Software Quality, Reliability, and Security (QRS)}}. \bibinfo{pages}{138--149}.
\newblock
\urldef\tempurl%
\url{https://doi.org/10.1109/QRS60937.2023.00023}
\showDOI{\tempurl}


\bibitem[\protect\citeauthoryear{Xie, Yang, He, and Chen}{Xie et~al\mbox{.}}{2021}]%
        {xie2021towards}
\bibfield{author}{\bibinfo{person}{Xiaoyuan Xie}, \bibinfo{person}{Haolin Yang}, \bibinfo{person}{Qiang He}, {and} \bibinfo{person}{Lin Chen}.} \bibinfo{year}{2021}\natexlab{}.
\newblock \showarticletitle{Towards Understanding Tool-chain Bugs in the LLVM Compiler Infrastructure}. In \bibinfo{booktitle}{\emph{2021 IEEE International Conference on Software Analysis, Evolution and Reengineering (SANER)}}. IEEE, \bibinfo{pages}{1--11}.
\newblock


\bibitem[\protect\citeauthoryear{Xiong, Xu, Su, Sun, Wang, Wen, Pu, He, and Su}{Xiong et~al\mbox{.}}{2023}]%
        {xiong2023an}
\bibfield{author}{\bibinfo{person}{Yiheng Xiong}, \bibinfo{person}{Mengqian Xu}, \bibinfo{person}{Ting Su}, \bibinfo{person}{Jingling Sun}, \bibinfo{person}{Jue Wang}, \bibinfo{person}{He Wen}, \bibinfo{person}{Geguang Pu}, \bibinfo{person}{Jifeng He}, {and} \bibinfo{person}{Zhendong Su}.} \bibinfo{year}{2023}\natexlab{}.
\newblock \showarticletitle{An Empirical Study of Functional Bugs in Android Apps}. In \bibinfo{booktitle}{\emph{Proceedings of the 32nd ACM SIGSOFT International Symposium on Software Testing and Analysis}} (Seattle, WA, USA) \emph{(\bibinfo{series}{ISSTA 2023})}. \bibinfo{publisher}{Association for Computing Machinery}, \bibinfo{address}{New York, NY, USA}, \bibinfo{pages}{1319–1331}.
\newblock
\showISBNx{9798400702211}
\urldef\tempurl%
\url{https://doi.org/10.1145/3597926.3598138}
\showDOI{\tempurl}


\bibitem[\protect\citeauthoryear{Xu, Chen, Sun, Zhou, and Lyu}{Xu et~al\mbox{.}}{2021}]%
        {xu2021memory}
\bibfield{author}{\bibinfo{person}{Hui Xu}, \bibinfo{person}{Zhuangbin Chen}, \bibinfo{person}{Mingshen Sun}, \bibinfo{person}{Yangfan Zhou}, {and} \bibinfo{person}{Michael~R Lyu}.} \bibinfo{year}{2021}\natexlab{}.
\newblock \showarticletitle{Memory-safety challenge considered solved? An in-depth study with all Rust CVEs}.
\newblock \bibinfo{journal}{\emph{ACM Transactions on Software Engineering and Methodology (TOSEM)}} \bibinfo{volume}{31}, \bibinfo{number}{1} (\bibinfo{year}{2021}), \bibinfo{pages}{1--25}.
\newblock


\bibitem[\protect\citeauthoryear{Yang, Gao, Liu, Li, and Xue}{Yang et~al\mbox{.}}{2024}]%
        {yang2024rusttwins}
\bibfield{author}{\bibinfo{person}{Wenzhang Yang}, \bibinfo{person}{Cuifeng Gao}, \bibinfo{person}{Xiaoyuan Liu}, \bibinfo{person}{Yuekang Li}, {and} \bibinfo{person}{Yinxing Xue}.} \bibinfo{year}{2024}\natexlab{}.
\newblock \showarticletitle{Rust-twins: Automatic Rust Compiler Testing through Program Mutation and Dual Macros Generation}. In \bibinfo{booktitle}{\emph{Proceedings of the 39th IEEE/ACM International Conference on Automated Software Engineering}} (Sacramento, CA, USA) \emph{(\bibinfo{series}{ASE '24})}. \bibinfo{publisher}{Association for Computing Machinery}, \bibinfo{address}{New York, NY, USA}, \bibinfo{pages}{631–642}.
\newblock
\showISBNx{9798400712487}
\urldef\tempurl%
\url{https://doi.org/10.1145/3691620.3695059}
\showDOI{\tempurl}


\bibitem[\protect\citeauthoryear{Yang, Chen, Eide, and Regehr}{Yang et~al\mbox{.}}{2011}]%
        {yang2011finding}
\bibfield{author}{\bibinfo{person}{Xuejun Yang}, \bibinfo{person}{Yang Chen}, \bibinfo{person}{Eric Eide}, {and} \bibinfo{person}{John Regehr}.} \bibinfo{year}{2011}\natexlab{}.
\newblock \showarticletitle{Finding and understanding bugs in C compilers}. In \bibinfo{booktitle}{\emph{Proceedings of the 32nd ACM SIGPLAN conference on Programming language design and implementation}}. \bibinfo{pages}{283--294}.
\newblock


\bibitem[\protect\citeauthoryear{Zhang, Kundu, Portokalidis, and Xu}{Zhang et~al\mbox{.}}{2023}]%
        {Zhang2023On}
\bibfield{author}{\bibinfo{person}{Yuchen Zhang}, \bibinfo{person}{Ashish Kundu}, \bibinfo{person}{Georgios Portokalidis}, {and} \bibinfo{person}{Jun Xu}.} \bibinfo{year}{2023}\natexlab{}.
\newblock \showarticletitle{On the Dual Nature of Necessity in Use of Rust Unsafe Code} \emph{(\bibinfo{series}{ESEC/FSE 2023})}. \bibinfo{publisher}{Association for Computing Machinery}, \bibinfo{address}{New York, NY, USA}, \bibinfo{pages}{2032–2037}.
\newblock
\showISBNx{9798400703270}
\urldef\tempurl%
\url{https://doi.org/10.1145/3611643.3613878}
\showDOI{\tempurl}


\bibitem[\protect\citeauthoryear{Zheng, Wan, Zhang, Chang, and Lo}{Zheng et~al\mbox{.}}{2023}]%
        {Zheng2023A}
\bibfield{author}{\bibinfo{person}{Xiaoye Zheng}, \bibinfo{person}{Zhiyuan Wan}, \bibinfo{person}{Yun Zhang}, \bibinfo{person}{Rui Chang}, {and} \bibinfo{person}{David Lo}.} \bibinfo{year}{2023}\natexlab{}.
\newblock \showarticletitle{A Closer Look at the Security Risks in the Rust Ecosystem}.
\newblock \bibinfo{journal}{\emph{ACM Trans. Softw. Eng. Methodol.}} \bibinfo{volume}{33}, \bibinfo{number}{2}, Article \bibinfo{articleno}{34} (\bibinfo{date}{dec} \bibinfo{year}{2023}), \bibinfo{numpages}{30}~pages.
\newblock
\showISSN{1049-331X}
\urldef\tempurl%
\url{https://doi.org/10.1145/3624738}
\showDOI{\tempurl}


\bibitem[\protect\citeauthoryear{Zhou, Ren, Gao, and Jiang}{Zhou et~al\mbox{.}}{2021}]%
        {zhou2021empirical}
\bibfield{author}{\bibinfo{person}{Zhide Zhou}, \bibinfo{person}{Zhilei Ren}, \bibinfo{person}{Guojun Gao}, {and} \bibinfo{person}{He Jiang}.} \bibinfo{year}{2021}\natexlab{}.
\newblock \showarticletitle{An empirical study of optimization bugs in GCC and LLVM}.
\newblock \bibinfo{journal}{\emph{Journal of Systems and Software}}  \bibinfo{volume}{174} (\bibinfo{year}{2021}), \bibinfo{pages}{110884}.
\newblock


\bibitem[\protect\citeauthoryear{Zhu, Zhang, Qin, Xiong, and Song}{Zhu et~al\mbox{.}}{2022}]%
        {zhu2022learning}
\bibfield{author}{\bibinfo{person}{Shuofei Zhu}, \bibinfo{person}{Ziyi Zhang}, \bibinfo{person}{Boqin Qin}, \bibinfo{person}{Aiping Xiong}, {and} \bibinfo{person}{Linhai Song}.} \bibinfo{year}{2022}\natexlab{}.
\newblock \showarticletitle{Learning and programming challenges of rust: A mixed-methods study}. In \bibinfo{booktitle}{\emph{Proceedings of the 44th International Conference on Software Engineering}}. \bibinfo{pages}{1269--1281}.
\newblock


\end{thebibliography}

\end{document}